\documentclass[A4paper,11pt]{article}
\usepackage{graphics,graphicx,epsfig,wrapfig}
\usepackage{longtable}  
\usepackage{amsmath,verbatim,amssymb,color,lscape}
\usepackage{kotex}
\usepackage{natbib}
\usepackage{lastpage}
\usepackage{multirow} 
\usepackage{makecell}
\usepackage{authblk}
\usepackage{bm}
\usepackage{indentfirst}
\usepackage[ruled,vlined]{algorithm2e}
\usepackage{algorithmic}
\usepackage{dirtytalk}
\usepackage[normalem]{ulem}
\usepackage[export]{adjustbox}
\usepackage{xcolor}
\usepackage{verbatim}
\usepackage{subcaption}
\usepackage{booktabs}
\usepackage{float}
\usepackage{url}
\usepackage{enumitem}
\usepackage{tikz}
\usetikzlibrary{shapes.geometric, arrows.meta, positioning, fit, backgrounds}

\usepackage{xr-hyper}
\newcommand{\hide}[1]{}

\makeatletter

\usepackage[outerbars,color]{changebar}
\ifx\pdfoutput\undefined
\else\ifnum\pdfoutput>0
  \usepackage{pdfcolmk}
\fi\fi
\cbcolor{black}

\usepackage[margin=1.0in]{geometry}
\usepackage{tikz}
\usetikzlibrary{bayesnet}
\usetikzlibrary{fit,positioning}

\newfont{\rmm}{cmr10 at 11pt}
\rmm

\title{
Constructing Reliable Social Networks from Conversational Data: An Ensemble Prompt Engineering Approach with Uncertainty Quantification}

\author[1]{Gwanghee Kim}
\author[1,2]{Ick Hoon Jin}
\author[3]{Minjeong Jeon}
\affil[1]{Department of Statistics and Data Science, Yonsei University. Republic of Korea.}
\affil[2]{Department of Applied Statistics, Yonsei University. Republic of Korea.}
\affil[3]{School of Education and Information Studies, University of California, Los Angeles. USA.}
\date{}

\begin{document}
\maketitle

\begin{abstract}
\noindent \textbf{Abstract}\\
Conversational data are central to the study of interaction dynamics and social structures across psychological research. However, constructing structured social networks from unstructured conversational data remains a major methodological challenge. This study presents a pipeline for \textcolor{black}{network construction} using prompt engineering. We employ an ensemble of \textcolor{black}{five} Large Language Models (LLMs) with majority voting to automate utterance classification, \textcolor{black}{reducing dependence on manual coding without task-specific parameter fine-tuning}. Classification \textcolor{black}{uncertainty} is assessed through an uncertainty quantification framework based on Shannon entropy, which \textcolor{black}{can be used to prioritize ambiguous cases for review}. The classified utterances are used to construct directed interaction networks for subsequent analysis. \textcolor{black}{Reliability and accuracy are established relative to the criterion-referenced and human comparisons reported here rather than as unconditional guarantees across settings.} We demonstrate the utility of this approach through two illustrative applications to classroom interaction data: network centrality analysis to characterize participant roles, and network mediation analysis using the additive and multiplicative effects network (AMEN) model to examine how interaction structures mediate the relationship between gender and mathematics performance. This pipeline provides a scalable foundation for automated network construction from conversational data across diverse research contexts.
\end{abstract}
\newpage

\section{Introduction}
\label{sec:intro}

Conversational data serve as a fundamental resource for investigating interaction dynamics and social structures across diverse psychological domains. Therapy transcripts in clinical psychology, for example, reveal therapeutic processes and treatment mechanisms \citep{imel2015computational, goldberg2020machine, abdou2025leveraging}, while social psychologists analyze group interactions to understand how conversational dynamics shape collective intelligence \citep{woolley2010evidence}. Similarly, organizational researchers examine team communications to identify patterns linked to team performance \citep{marlow2018team}, and educational psychologists study classroom dialogues to uncover the mechanisms through which peer discourse supports learning \citep{howe2013classroom}.

Among these settings, classroom dialogue offers a particularly valuable context for methodological development. Classroom interactions exhibit the full complexity of multi-party conversations---diverse utterance types, frequent topic shifts, and context-dependent meanings \citep{gardner2019classroom}---while also providing external performance criteria against which the utility of derived network measures can be evaluated. Empirical research has consistently demonstrated that network properties of classroom interactions are associated with measurable student outcomes: \citet{sedova2019those} found correlations between dialogue participation patterns and academic performance and \citet{mercer2010analysis} showed that the structure of student conversations fosters collaborative problem-solving skills. This combination makes classroom dialogue a practical {\color{black}setting in which to develop and examine the proposed network-construction procedure. Accordingly, the present study applies the full pipeline to the classroom corpus of \citet{webb2023learning}, which comprises 20 third-grade students observed across six mathematics lessons.}

\paragraph*{Conventional Approaches}

Despite its analytical promise, classroom conversation data present substantial practical challenges. Beyond structural complexities, these conversations encompass speaker characteristics (e.g., individual speaking style, prior experience), emotional tone (e.g., enthusiasm, uncertainty, affect), and contextual factors (e.g., task type, classroom environment), making them difficult to analyze manually.
 
To address such complexity, researchers often simplify the analysis by categorizing the intent of individual messages or utterances using predetermined coding schemes. For example, researchers may classify utterances according to function, such as whether they are questions, answers, elaborations, or agreements \citep{webb2014engaging, franke2015student}. This strategy entails classifying utterances based on semantic, syntactic, and pragmatic features, and using these classifications to gain insights into specific aspects of the interaction.

However, traditional analytical methods rely heavily on manual labeling, requiring trained human coders to analyze conversational data, which is both labor-intensive and time-consuming. For example, \citet{chi1997quantifying} described a protocol for \say{verbal analysis}, involving a detailed examination of transcripts by trained human coders to identify a series of predefined categories. \citet{de2006content} discuss the application of \say{coding schemes}, which are also applied manually by human coders to classify utterances in online forums. Manual annotation processes often require extensive training to ensure reliability, which can be costly to implement on a large scale. 

These limitations highlight the need for automated analytical strategies to thoroughly examine such conversational data. Although several deep learning methods have been proposed for automatic dialogue classification \citep{liu2017dadnn, CASA, song2021automatic}, they depend on large labeled datasets that are difficult to acquire and may not generalize well across different research objectives. For example, \citet{CASA} proposed a context-aware self-attention model for general dialogue act classification, but it requires extensive manual annotation for adaptation to specific tasks. On the other hand, \citet{song2021automatic} proposed an automated model for classroom dialogue classification by combining a \textcolor{black}{convolutional neural network (CNN) and a bidirectional long short-term memory (Bi-LSTM)}, but it also requires large labeled datasets and may have difficulty generalizing to diverse research objectives. These challenges motivate the development of alternative approaches that minimize reliance on pre-labeled datasets and offer greater flexibility.

\paragraph*{Prompt Engineering} 

Large language models like Generative Pre-trained Transformer \citep[GPT]{gpt1, gpt2, gpt3} offer an effective alternative that overcomes these limitations. Through training on extensive text datasets, GPT demonstrates remarkable versatility in executing diverse tasks with minimal task-specific training. A distinctive characteristic of GPT is its capacity for multitask performance using a single pre-trained model. The performance of GPT can be further enhanced through carefully constructed \say{prompts} that provide explicit instructions or contextual guidance. This approach, known as prompt engineering, can significantly improve model outputs \citep{liu2023pre}. Moreover, including a few examples within the prompt enables accurate output generation without requiring updates to model parameters \citep{gpt3}. In sentiment analysis, for instance, incorporating directive statements such as \say{Classify the sentiment of this sentence as positive, negative, or neutral} along with examples like \say{Example: Sentence: I love this product. Sentiment: Positive} within the prompt helps to guide the model explicitly and improves classification accuracy.

Building on these capabilities, recent studies have explored prompt engineering for automating classroom dialogue analysis. \citet{long2024evaluating} reported that LLMs like GPT-4 can perform coding tasks up to 30 times faster than human researchers while achieving over 90\% agreement for relatively straightforward classification tasks. On the other hand, \citet{ganesh2024prompting} found that while models perform well on tasks with clear linguistic cues, they exhibit unstable and inconsistent performance on complex, theoretically motivated qualitative coding tasks requiring nuanced interpretation.{\color{black}\citet{ganesh2024prompting} also found that fine-tuned models can be more reliable when sufficient scheme-specific labeled data are available. However, constructing such a training corpus and updating model parameters impose precisely the data and adaptation requirements that motivate a prompt-based alternative.}

{\color{black}Accordingly, the present study does not frame prompt-based coding as an attempt to reproduce fine-tuning without labeled data. Instead, it combines an equal-weight ensemble with explicit uncertainty diagnostics without requiring task-specific parameter updates.}

\paragraph*{From Utterances to Interaction Networks}

While reliable utterance classification produces a structured record of conversational events, treating each classified utterance in isolation misses the relational structure among participants that gives rise to broader social dynamics \citep{bruun2019network}. A natural next step, therefore, is to represent interactions as networks, where participants serve as nodes and communicative exchanges serve as edges. {\color{black}Unlike ties that are observed directly in many conventional social networks, discourse relations must be inferred from the coded function of an utterance together with its local conversational context and reference target. In the present application, EXP and EOI serve as primary indicators of relation type and direction and the EOI engagement level serves as a secondary indicator of edge weight. Related distinctions between contributing ideas and engaging with the reasoning of others appear in Accountable Talk, ICAP, SEDA and exploratory talk frameworks \citep{michaels2008accountable, chi2014icap, hennessy2016seda, mercer1996quality}. Other discourse schemes require scheme-specific manual-grounded prompts and explicit construction rules.} This representation enables explicit measurement of relational properties such as degree centrality, reciprocity, and network centralization \citep{liyanage2021student, dawson2008study}. Prior work has shown that such metrics are substantively informative: for instance, centrality correlates with perceived sense of community \citep{dawson2008study}, and network structure captures meaningful differences in interaction quality across learning groups \citep{liyanage2021student}. These findings motivate the approach taken in the present study, which transforms classified dialogue data into directed interaction networks as the basis for subsequent quantitative analysis.

{\color{black}In this paper, we present a methodology with two stages. Manual-grounded automated coding transforms utterances into structured conversational events and prespecified construction rules transform those events into directed weighted networks. Network analysis follows as a separate downstream step and is not used to justify the construction rules. For the coding stage, we aggregate five instruction-tuned LLMs using majority voting. All five models receive equal vote weight, so the ensemble records cross-model agreement without assigning a priori priority to a particular model \citep{yue2024large, jiang2023llmblender, lu2024merge, chen2025harnessing}. We further quantify disagreement using ensemble-vote entropy and, where model logits are available, intra-model entropy. These diagnostics identify utterances that can be prioritized for review. The present study did not implement expert adjudication of the flagged utterances in the classroom corpus and uncertainty-based triage alone does not establish classification accuracy.}

Finally, to demonstrate the empirical utility of the constructed networks, we present two complementary illustrative applications using a real-world classroom dataset. First, we employ network centrality analysis to characterize interaction patterns and identify structurally prominent participants in classroom discourse. Second, we apply network mediation analysis using the additive and multiplicative effects network (AMEN) model \citep{amen} with a negative binomial specification to examine how classroom interaction networks mediate the relationship between gender and mathematics performance. These applications are intended as demonstrations of the methodology's capacity to support diverse analytical approaches---both descriptive and inferential---rather than as definitive substantive conclusions; researchers may apply any network analysis method appropriate to their own research questions.

The remainder of this paper is structured to present our methodology and findings systematically. Section 2 examines our prompt engineering approach, detailing the automated utterance classification process, the associated validation framework, and the creation of labeled interaction data. Section 3 demonstrates the analytical utility of our methodology through two illustrative applications: network centrality analysis for descriptive characterization of interaction patterns, and network mediation analysis for statistical modeling of relationships between individual attributes and outcomes. Section 4 concludes by exploring the broader implications of our research and identifying promising directions for future studies that extend this methodological approach.

\section{Prompt Engineering for Utterance Classification}
\label{sec:prompt}




\tikzset{
  block/.style      = {rectangle, rounded corners=4pt, draw=black!70, fill=white,
                       text width=5.0cm, align=center, minimum height=0.80cm,
                       font=\small},
  valblock/.style   = {rectangle, rounded corners=4pt, draw=black!50,
                       fill=gray!10, text width=4.4cm, align=center,
                       minimum height=0.70cm, font=\small\itshape},
  appblock/.style   = {rectangle, rounded corners=4pt, draw=black!40,
                       dashed, fill=white, text width=5.0cm, align=center,
                       minimum height=0.80cm, font=\small},
  decision/.style   = {diamond, draw=black!70, fill=white, aspect=3.4,
                       text width=3.4cm, align=center, font=\small,
                       inner sep=2pt},
  arrow/.style      = {-{Stealth[length=5pt]}, thick},
  thinnarrow/.style = {-{Stealth[length=4pt]}, thin, draw=black!55},
  groupbox/.style   = {draw=black!40, rounded corners=6pt, dashed,
                       fill=gray!5, inner sep=10pt},
}

\begin{figure}[!ht]
\centering
\begin{tikzpicture}[
  node distance=0.8cm and 3.0cm,
  >=Stealth,
  thick,
  block/.style={
    rectangle, rounded corners=4pt, draw=black!80, fill=white,
    text width=5.2cm, align=center, minimum height=1.0cm, font=\small
  },
  valblock/.style={
    rectangle, rounded corners=4pt, draw=black!60, fill=gray!5,
    text width=5.8cm, align=center, minimum height=0.75cm, font=\small\itshape
  },
  appblock/.style={
    rectangle, rounded corners=4pt, draw=black!80, fill=white, dashed,
    text width=5.2cm, align=center, minimum height=1.0cm, font=\small
  },
  arrow/.style={-{Stealth[length=6pt]}, thick},
  thinnarrow/.style={-{Stealth[length=6pt]}, thin, draw=black!60},
  groupbox/.style={
    draw=black!30, rounded corners=8pt, fill=gray!2, inner sep=15pt
  }
]

  \node[block] (data)      {Conversational Data};
  \node[block, below=of data] (prompt) {Prompt Engineering\\{\footnotesize Chain-of-Thought \& Few-Shot}};
  \node[block, below=of prompt] (ensemble) {Ensemble Classification\\{\footnotesize \textcolor{black}{Five} LLMs \& Majority Voting}};
  \node[block, below=1.2cm of ensemble] (network) {Network Construction};
  \node[appblock, below=0.8cm of network] (analysis) {Network Analysis};

  \node[valblock, right=1.6cm of ensemble, yshift=0.7cm] (v1) {Inter-Model Consistency\\{\footnotesize (Cohen's $\kappa$, Fleiss' $\kappa$)}};
  \node[valblock, below=0.4cm of v1] (v2) {Uncertainty Quantification\\{\footnotesize (Shannon Entropy)}\\[3pt]{\footnotesize Inter-Model: Ensemble Votes}\\{\footnotesize Intra-Model: Sequence \& Token Logits}};

  \begin{scope}[on background layer]
    \node[groupbox, fit=(v1)(v2),
          label={[font=\small\bfseries, yshift=2pt]above:Validation Framework}] (vbox) {};
  \end{scope}

  \draw[arrow] (data) -- (prompt);
  \draw[arrow] (prompt) -- (ensemble);
  \draw[arrow] (ensemble) -- (network);
  \draw[arrow, dashed] (network) -- (analysis);

  \draw[thinnarrow] (ensemble.east) -- (vbox.west |- ensemble.east);

\end{tikzpicture}
  \caption{
  Overview of the proposed methodology. The main pipeline (left) transforms conversational data into interaction networks through prompt engineering, ensemble classification via \textcolor{black}{five} LLMs with majority voting, and network construction. A parallel validation framework (right) assesses classification reliability using inter-model consistency metrics (Cohen's $\kappa$, Fleiss' $\kappa$) and quantifies both inter- and intra-model uncertainty via Shannon entropy. The constructed networks serve as input for downstream analyses (bottom), illustrated here through centrality and mediation applications.
  }
\label{fig:flowchart}
\end{figure}

\subsection{Classroom Interaction Data and Coding Scheme}

We analyze classroom interaction data from \citet{webb2023learning}, which includes observations of 20 third-grade students from a classroom located in a low-income urban community in Southern California. Researchers collected data over a five-month period, during which time they recorded six one-hour math lessons using strategically placed cameras to capture comprehensive classroom interactions between teachers and students, as well as interactions among students. The recordings were carefully transcribed to include contextual information, including speaker identities, timestamps, and task descriptions. 

{\color{black}To analyze the dialogue transcripts, we operationalize the two contribution types defined by \citet{webb2014engaging}, explaining own ideas (EXP) and engaging with the ideas of others (EOI), together with a residual uncorrelated (UC) category that we add for utterances with no mathematical contribution, yielding three primary categories. EXP identifies utterances that articulate mathematical reasoning, strategies or justification. EOI identifies responses that engage with the contribution of another participant and is further differentiated into low engagement (a brief acknowledgment or agreement), medium engagement (a specific reference or clarification request) and high engagement (new information, a reasoned challenge or an extension of the referenced idea). UC covers turns that supply neither an explanation nor a qualifying engagement with another idea, including procedural, off-topic or otherwise non-analytic utterances.}

{\color{black} A central challenge in analyzing dialogue data arises from the context-dependent nature of utterances. The meaning of conversations often depends on the surrounding dialogue. The following excerpt from the classroom corpus illustrates how the coding uses the preceding contribution and the reference target:}
\begin{quote}\small
\textcolor{black}{\textbf{Esther}: And there is 1 here (points to her paper). So that means 100 (starts to erase something).\\
\textbf{Mya}: It's going to be 400, right?\\
\textbf{Esther}: One hundred forty-four. Because we did not add the one. So, we have to put it right here points to her paper. And then it equals 144.}
\end{quote}
\noindent {\color{black}The first turn of Esther is coded as EXP because it explains her representation. The question from Mya is coded as medium EOI directed to Esther because it requests clarification of that representation. The final turn of Esther is coded as high EOI directed to Mya because it corrects the proposed result by supplying a mathematical reason. This sequence illustrates the context-dependent application of the coding rules.}

\subsection{Prompt Engineering Methodology}

The inherent complexity and contextual richness of classroom interactions require a robust method to classify and analyze the intent of dialogue. As discussed in Section \ref{sec:intro}, existing analytical approaches require manual labeling, which is resource-intensive, while deep learning methods require extensive labeled datasets. Our methodology addresses these issues through prompt engineering, by leveraging pre-trained models to \textcolor{black}{classify dialogue intents without task-specific parameter fine-tuning or a large labeled training corpus}. This approach maximizes the capabilities of pre-trained models while maintaining efficiency in resource utilization and processing time.

In-context learning (ICL) is a specialized application of prompt engineering in which pre-trained LLMs learn exclusively from information provided within the prompt, eliminating the need for parameter updates. This approach parallels human learning processes through inference while minimizing computational requirements for task adaptation \citep{dong2024survey}. Few-shot prompting, one of the ICL strategies, enhances model comprehension by incorporating targeted task examples within the prompt. The effectiveness of few-shot prompts in ICL has been explained through probabilistic frameworks such as hidden Markov models \citep{baum1966statistical}, where carefully selected examples guide model reasoning by establishing sequential dependencies that reduce prediction uncertainty \citep{xie2021explanation}. \citet{min2022rethinking} further demonstrates that optimal performance depends not merely on increasing the \textcolor{black}{number} of examples, but on developing prompt formats that precisely align with the specific requirements of the task, including both structural elements and linguistic formulation.

In our study, the prompts are systematically designed using a \textcolor{black}{two-stage structured classification-and-rationale procedure to operationalize the coding criteria consistently}. The first step generates explicit reasoning for the classification decision, while the second step applies this reasoning to produce the actual categorization. The system prompt incorporates detailed classification criteria and categorization rules that define the utterance classification task. Specifically, it provides explicit guidelines for categorizing interactions as ``explaining own ideas,'' ``engaging with others' ideas,'' \textcolor{black}{or ``uncorrelated,''} with the \textcolor{black}{EOI category} further differentiated into low, medium, or high engagement levels. In addition, the system prompt outlines specific criteria for making classification decisions, including handling special cases and exceptions. \textcolor{black}{Detailed reasoning examples with classifications clarify how the definitions and category boundaries apply, and the response format fixes the primary category, engagement level where applicable, reference target, and concise rationale.}

The user prompt complements this framework by providing essential contextual information required for classification decisions, including previous utterances, speaker information, and \textcolor{black}{candidate reference information} that enable the model to understand the flow of conversation and the relationships between participants. This contextual grounding facilitates more nuanced classification of utterances into the three main categories: \textit{explain your own ideas}, \textit{engage with others' ideas}, or \textit{uncorrelated}. The prompts incorporate \textcolor{black}{detailed reasoning examples with classifications} that clarify distinctions between categories and demonstrate practical application of classification rules. \textcolor{black}{These examples were constructed to fix the intended output format, a brief rationale followed by the classification decision, rather than selected as a representative sample of the corpus.} \textcolor{black}{For the dataset from \cite{webb2023learning}, where no gold standard exists, model outputs were inspected to align the coding intention with the operational instructions, after which the prompt was fixed.} Detailed prompts are provided in Section \ref{supp-sec:prompt} of the Supplementary Material. \textcolor{black}{To reduce response variability, we set the temperature parameter to zero.}

{\color{black}While careful prompt design improves classification quality for a single model, individual LLMs remain susceptible to model-specific variation on ambiguous utterances. We therefore employ an ensemble of five instruction-tuned LLMs comprising Claude Sonnet 5 \citep{anthropic2026sonnet5}, GPT-5.6 Terra \citep{openai2026gpt56}, Gemini-3.8 Flash \citep{googledeepmind2026gemini38}, Gemma4-31B-it \citep{gemmateam2026gemma4}, and Qwen3.6-27B \citep{qwenteam2026qwen36}. The first three were accessed through commercial APIs via OpenRouter, and the latter two were deployed locally using open-weight checkpoints. All five models receive equal voting weight, and the most frequent classification is assigned as the final label through majority voting. If all five primary categories are EOI but engagement levels are tied, the lower engagement level is selected conservatively. Remaining ties are resolved by the three commercial-model votes and, when required, their prespecified order. This rule provides a symmetric and reproducible aggregation procedure when no single model can be assumed to dominate across utterances, without implying that the rule is optimally weighted. Detailed specifications are provided in the Supplementary Material.}

All models are prompted using \textcolor{black}{the same coding definitions, contextual fields, and reasoning examples with classifications}. Prompt construction adheres to the \textcolor{black}{appropriate message format for each model interface} to ensure methodological consistency and reproducibility across the entire ensemble. Detailed model descriptions, \textcolor{black}{access and execution records}, and formatting guidelines are provided in Section \ref{supp-sec:llm} of the Supplementary Material. This standardized prompting approach \textcolor{black}{mitigates the influence of model-specific formatting differences} on classification outcomes.

\subsection{Utterance Classification Results}

The classification process implements a \textcolor{black}{two-stage structured classification-and-rationale procedure}, requiring models to first generate explicit reasoning before producing the final categorization. Each \textcolor{black}{reasoning example with a classification} demonstrates explicit reasoning processes covering speaker identification, utterance content analysis, classification labeling, engagement level assessment, and relevant contextual references. Complete prompt specifications, including these examples, are provided in Section \ref{supp-sec:prompt} of the Supplementary Material. \textcolor{black}{The two open-weight models were executed locally on two NVIDIA RTX A6000 GPUs, whereas the three commercial models were accessed through their APIs. Execution details and token usage are reported in the Supplementary Material.}

Following the utterance classification, we excluded entries labeled as ``uncorrelated'' from subsequent \textcolor{black}{network} analyses. The remaining utterances, classified as EXP and EOI, were segregated to construct distinct interaction networks for each type of discourse. As a result, a total of \textcolor{black}{4,235} utterances were \textcolor{black}{classified}, of which \textcolor{black}{440 were classified as EXP, 670 as EOI, and 3,125 as uncorrelated}. The \textcolor{black}{EOI} category was further divided into three subcategories based on engagement level: \textit{high}, \textit{medium}, and \textit{low}. Table \ref{tab:classification} presents a detailed breakdown of these classifications.

\begin{table}[H]
\centering
\small
\setlength{\tabcolsep}{7pt}
\renewcommand{\arraystretch}{1.08}
\begin{tabular}{llc}
\Xhline{2\arrayrulewidth}
\textbf{Utterance Type} & \textbf{Engagement Level} & \textbf{Number of Utterances} \\
\hline
Explain Own Idea (EXP)  & & \textcolor{black}{440} \\
Engage Others Idea (EOI) & Total & \textcolor{black}{670} \\
 & High Engagement & \textcolor{black}{124} \\
 & Medium Engagement & \textcolor{black}{187} \\
 & Low Engagement & \textcolor{black}{359} \\
Uncorrelated  & & \textcolor{black}{3,125}\\
\hline
Total Utterances & & \textcolor{black}{4,235} \\
\Xhline{2\arrayrulewidth}
\end{tabular}
\caption{Classification Results by Utterance Type and Engagement Level}
\label{tab:classification}
\end{table}

We provide three illustrative examples to demonstrate how the classification scheme operates in practice across diverse interaction types. \textcolor{black}{These examples are not presented as a validation sample.} Table \ref{tab:example1} demonstrates how the system effectively classifies instances in which students explain their own ideas, including responses to the questions. Student explanations of mathematical reasoning, often presented in extended detail, are labeled \textit{explaining their own idea}. In contrast, brief affirmations or interjections that do not reflect substantial communicative intent are correctly classified as \textit{uncorrelated}. The example also illustrates the medium-level engagement classification, particularly when students pose clarifying questions such as \say{How do you know it was 35?} about others' ideas. \textcolor{black}{Kimberly's later counting-and-turn-management utterance is classified as uncorrelated because it does not add a conceptual explanation or qualifying engagement with another idea.}

\clearpage
\begin{table}[H]
    \centering
    \small{
    \setlength{\tabcolsep}{4pt}
    \renewcommand{\arraystretch}{1.0}
    \begin{tabular}{>{\raggedleft\arraybackslash}m{1.5cm}|
                    >{\raggedright\arraybackslash}m{7.5cm}|
                    >{\centering\arraybackslash}m{3cm}|
                    >{\centering\arraybackslash}m{3cm}}
    \Xhline{2\arrayrulewidth}
    \textbf{Name} & \textbf{Utterances} & \textbf{Label} & \textbf{Engagement Level}\\
    \Xhline{2\arrayrulewidth}
    Kimberly & I thought you did just it by 5 because like (points to N's notebook). & engage others idea & medium\\
    \hline
    Natalie & Then I got to 5s and made 10, so I put 5 and 5 to (inaudible). And then I counted how much there was both and…it was 35. & explain own idea & -\\
    \hline
    Kimberly & Okay. Samantha go next. & uncorrelated & -\\
    \hline
    Samantha & What I did was right here. I did yesterday and then I didn't have enough for everyone. And then I passed the last 10 because if I put it in with these, wouldn't it be equal and then I split it. And then I carried it in and then I divided by once, and then I counted it and then I got 1, 2, 3, 4, 5 and this was 30, and then I got 35. & explain own idea & -\\
    \hline
    Kimberly & What I did was like I did like Samantha's but I just labeled one day and two day, and then I just made a cross. Then, here I did the same thing, I got the seven 10s, but when I end up with 2, I just cross 2 off and then the 10. But then I noticed this was like that so I turned them into little 10 ones. And then I got 1, 2, 3, 4, 5, 6, 7, 8, 9. 10. I got 35. & explain own idea & -\\
    \hline
    Natalie & (has a look of disbelief) How do you know it was 35? & engage others idea & medium\\
    \hline
    Kimberly & Like I counted it. 10, 20. I already knew this was 20, 30 & explain own idea & -\\
    \hline
    Natalie & 40. & uncorrelated & -\\
    \hline
    Kimberly & (laughs)…30, 31, 32, 33, 34, 35. And that's my answer for how I got to 35. (adjusts camera) And now can I go first for the second number share? & \textcolor{black}{uncorrelated} & -\\
    \hline
    Natalie & Yes. & uncorrelated & -\\
    \Xhline{2\arrayrulewidth}
    \end{tabular}
    }
    \caption{Examples: Identifying labels (explain vs. engage).}
    \label{tab:example1}
\end{table}

\clearpage
\begin{table}[H]
    \centering
    \small{\color{black}
    \setlength{\tabcolsep}{4pt}
    \renewcommand{\arraystretch}{1.0}
    \begin{tabular}{>{\raggedleft\arraybackslash}m{1.5cm}|
                    >{\raggedright\arraybackslash}m{7.5cm}|
                    >{\centering\arraybackslash}m{3cm}|
                    >{\centering\arraybackslash}m{3cm}}
    \Xhline{2\arrayrulewidth}
    \textbf{Name} & \textbf{Utterances} & \textbf{Label} & \textbf{Engagement Level}\\
    \Xhline{2\arrayrulewidth}
    Melissa & Why are we putting 15 in it? & engage others idea & medium\\
    \hline
    Natalie & Because we are gonna split it in 5, 5, and 5. & explain own idea & --\\
    \hline
    Melissa & Why are we gonna split it? & engage others idea & medium\\
    \hline
    Natalie & Because it's easier to do 5s instead of 15s. & explain own idea & --\\
    \hline
    Melissa & Isn't it more efficient to do 15? & engage others idea & high\\
    \hline
    Natalie & Well, in 5's you can just keep on counting because 5's are really easy. & engage others idea & high\\
    \hline
    Melissa & But don't you already know that 15 plus 15 is 30? So it's more easier than that way. & engage others idea & high\\
    \Xhline{2\arrayrulewidth}
    \end{tabular}
    }
    \caption{High-Level Engagement in Problem-Solving}
    \label{tab:example2}
\end{table}

\clearpage
{\color{black}Table \ref{tab:example2} presents a separate Natalie--Melissa sequence in which Natalie explains her strategy, Melissa asks specific clarification questions coded as medium EOI, and the students then propose and defend alternative strategies with mathematical reasons coded as high EOI. Melissa's EOI turns refer to Natalie, while Natalie's high-EOI response refers to Melissa.}

Table \ref{tab:example3} demonstrates the system's capacity to distinguish between mathematical discourse and off-topic conversations within classroom settings. While off-topic remarks are classified as \textit{uncorrelated}, mathematical explanations receive relevant classifications such as \textit{engage with others' ideas}. The example further illustrates how the system recognizes varying levels of engagement with peer contributions, identifying both medium- and low-level engagement instances. Subsequently, when students transition to explaining their own reasoning, the system correctly adjusts the classification accordingly. This sequence exemplifies the system's ability to maintain classification accuracy across diverse types of classroom discourse while maintaining focus on mathematically relevant interactions. \textcolor{black}{Kimberly's final procedural remark is classified as uncorrelated.}

\clearpage
\begin{table}[H]
    \centering
    \small{
    \setlength{\tabcolsep}{4pt}
\renewcommand{\arraystretch}{1.0} 
    \begin{tabular}{>{\raggedleft\arraybackslash}m{1.5cm}|
                    >{\raggedright\arraybackslash}m{7.5cm}|
                    >{\centering\arraybackslash}m{3cm}|
                    >{\centering\arraybackslash}m{3cm}}
    \Xhline{2\arrayrulewidth}
    \textbf{Name} & \textbf{Utterances} & \textbf{Label} & \textbf{Engagement Level}\\
    \Xhline{2\arrayrulewidth}
Natalie & Okay, let me go. & uncorrelated & -\\
\hline
Kimberly & I missed a number. & uncorrelated & -\\
\hline
Natalie & I'm looking at the hundreds chart. (refers to yellow sheet) And then I went to 75. & uncorrelated & -\\
\hline
Kimberly & Why did you go to 75? & engage others idea & medium\\
\hline
Natalie & Because that's how much shells… & explain own idea & -\\
\hline
Kimberly & (talks over N) No, it was 70. & engage others idea & low\\
\hline
Natalie & …that was how much shells she found. & explain own idea & -\\
\hline
Kimberly & It was 70. & uncorrelated & -\\
\hline
Natalie & On the second strategy. & uncorrelated & -\\
\hline
Kimberly & Oh! & uncorrelated & -\\
\hline
Natalie & And then, I was thinking how could I do it. Then, I was thinking about money, I don't know why. Then, I remembered a quarter, so then I know what that is. (pause) And then, I started like because I remember the quarters, so I went 20, 25, no… & explain own idea & -\\
\hline
Kimberly & (talking to another group) You should keep on sharing strategies. That's what we are doing. & uncorrelated & -\\
\hline
Samantha & We are already done. & uncorrelated & -\\
\hline
Kimberly & We are at the strategies, but she and me realized that we strategies on the 200 chart. & \textcolor{black}{uncorrelated} & -\\
\Xhline{2\arrayrulewidth}
\end{tabular}
}
    \caption{Differentiating Relevant and Off-Topic Remarks.}
    \label{tab:example3}
\end{table}

\clearpage
\subsection{Validation of Utterance Classification} 

To evaluate \textcolor{black}{multiple forms of evidence concerning} our utterance classification framework, we implemented a multi-faceted validation approach. This validation strategy includes \textcolor{black}{four} primary analytical components: inter-model consistency analysis, \textcolor{black}{external criterion evaluation}, human annotation comparison, and uncertainty quantification analysis. The inter-model consistency analysis assesses agreement levels between different LLMs, while human annotation comparison examines the consistency between human judgments and LLM-generated classifications. Additionally, we conducted both inter-model and intra-model uncertainty analyses to systematically identify instances \textcolor{black}{that can be prioritized for review} and to examine individual model confidence levels. These complementary approaches provide evidence about distinct aspects of classification performance. \textcolor{black}{None alone establishes full-corpus classification accuracy.}

\subsubsection{Inter-Model Consistency Analysis} 

We conducted an inter-model consistency analysis to evaluate the degree of agreement between LLMs in their classification decisions. This analysis employed two established agreement metrics that provide complementary perspectives on model alignment: Cohen's $\kappa$ \citep{cohen} and Fleiss' $\kappa$ \citep{fleiss}.  

\paragraph*{Cohen's $\kappa$} 

We utilized Cohen's $\kappa$ coefficient \citep{cohen} to assess pairwise agreement between individual model pairs, providing a chance-corrected measure of concordance for each model combination. Cohen's $\kappa$ is defined as:
\begin{equation}
\kappa_{Cohen} = \frac{p_o - p_e}{1 - p_e}
\end{equation}
where $p_o$ represents the observed proportional agreement between the two models under comparison, while $p_e$ denotes the expected proportional agreement that would occur due to chance alone. The metric provides a chance-corrected measure of agreement that operates within a standardized range from -1, indicating perfect disagreement, to +1, representing perfect agreement. A coefficient value of 0 means that the observed agreement is equivalent to what would be expected through random chance alone. The interpretation of Cohen's $\kappa$ values follows established conventions as originally proposed by \cite{cohenkappa}, with specific thresholds and their corresponding agreement levels detailed in Table \ref{tab:cohen_kappa_interpret}.

\begin{table}[htbp]
    \centering
    \begin{tabular}{cc}
    \hline
    \textbf{\boldsymbol{$\kappa$} Statistic} & \textbf{Strength of Agreement} \\
    \hline    
    $<$ 0.00 & Poor \\
    0.00 - 0.20 & Slight \\
    0.21 - 0.40 & Fair \\
    0.41 - 0.60 & Moderate \\
    0.61 - 0.80 & Substantial \\
    0.81 - 1.00 & Almost perfect \\
    \hline
    \end{tabular}
    \caption{    
    Standardized interpretation framework for Cohen's $\kappa$ coefficient values, providing qualitative assessments of inter-rater agreement strength.
    }

    \label{tab:cohen_kappa_interpret}
\end{table}

\begin{figure}[!ht]
    \centering
    \includegraphics[width=.7\textwidth]{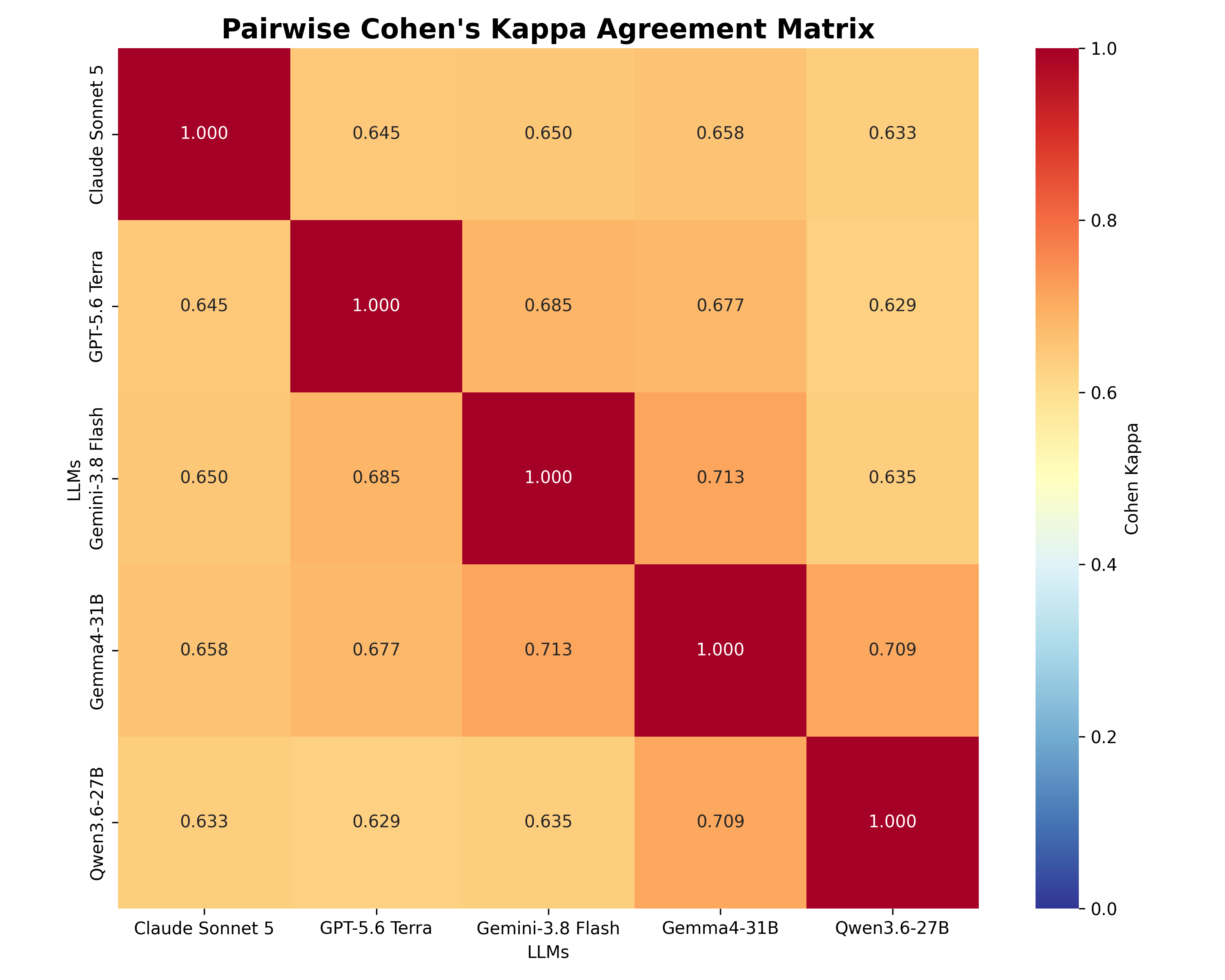}
    \caption{
    Heatmap visualization displaying pairwise Cohen's $\kappa$ coefficients computed across all possible combinations of the \textcolor{black}{five} LLMs included in our ensemble framework. Each cell quantifies the chance-corrected agreement between two specific models, with color intensity representing the magnitude of agreement strength. Darker color intensities correspond to higher agreement levels, while lighter shades indicate lower consensus between model pairs.
    }
    \label{fig:pairwise_cohen_llm}
\end{figure}

{\color{black}Figure \ref{fig:pairwise_cohen_llm} presents pairwise Cohen's $\kappa$ coefficients for the five ensemble models. The coefficients range from .6291 to .7134, with the highest agreement between Gemini-3.8 Flash and Gemma4-31B and the lowest between GPT-5.6 Terra and Qwen3.6-27B. All ten pairwise coefficients fall in the conventionally defined substantial-agreement range.}

\paragraph*{Fleiss' $\kappa$}

Additionally, we employed Fleiss' $\kappa$ \citep{fleiss} to evaluate multi-rater agreement across \textcolor{black}{the five models and obtain a measure of overall ensemble consistency}. Specifically, Fleiss' $\kappa$ extends Cohen's $\kappa$ framework to accommodate multiple raters in a single comprehensive analysis. Fleiss' $\kappa$ is defined as:
\begin{equation}
    \kappa_{Fleiss} = \frac{\bar{P} - \bar{P_e}}{1 - \bar{P_e}}
\end{equation}
where $\bar{P}$ represents the mean observed agreement across all classification items, while $\bar{P_e}$ denotes the mean expected agreement that would occur under the statistical assumption of independence among all \textcolor{black}{five} models. This metric captures ensemble-wide consistency patterns that cannot be detected through pairwise comparisons alone, providing \textcolor{black}{information about the overall coherence of the model outputs}.

\begin{table}[htbp]
    \centering
    \begin{tabular}{llc}
    \hline
    \textcolor{black}{\textbf{Group}} & \textcolor{black}{\textbf{Statistic}} & \textcolor{black}{\textbf{\boldsymbol{$\kappa$}}} \\
    \hline    
    \textcolor{black}{Five-model ensemble} & \textcolor{black}{Fleiss' $\kappa$} & \textcolor{black}{.7164} \\
    \textcolor{black}{Commercial API models (three)} & \textcolor{black}{Fleiss' $\kappa$} & \textcolor{black}{.7135} \\
    \textcolor{black}{Open-weight models (two)} & \textcolor{black}{Cohen's $\kappa$} & \textcolor{black}{.7092} \\
    \hline
    \end{tabular}
    \caption{
    \textcolor{black}{Chance-corrected agreement for the complete ensemble and model-source groupings. Cohen's $\kappa$ is reported for the two-model open-weight pair. Fleiss' $\kappa$ is reported for groups containing three or more models.}
    } 
    \label{tab:llm_fleiss_kappa}
\end{table}

{\color{black}The complete five-model ensemble achieved Fleiss' $\kappa=.7164$ across the 4,235 utterances. Agreement was similar for the three commercial API models (Fleiss' $\kappa=.7135$) and the two open-weight models (Cohen's $\kappa=.7092$). These coefficients quantify reproducibility across model outputs. They are not interpreted as classification accuracy in the absence of gold labels for the full Webb corpus.}

\subsubsection{\textcolor{black}{External Criterion Evaluation on TalkMoves}}

{\color{black}We additionally evaluated the same five-model ensemble procedure on TalkMoves, a separately annotated corpus of K--12 mathematics classroom discourse \citep{suresh2022talkmoves}. TalkMoves uses its own five student talk-move categories grounded in Accountable Talk. These categories are substantively related to explaining and engaging with ideas but are not treated as identical to Webb EXP, EOI, or the EOI sub-levels. Because a small reference-labeled pilot informed instruction and example alignment, we describe the following as development-informed, full-corpus criterion-referenced performance rather than an independent test-set estimate.}

{\color{black}Label distributions across all 63 TalkMoves episodes were reviewed, and whole episodes were selected to provide the most balanced available coverage of the five labels, especially the rarer peer-engagement categories. A 234-utterance development pilot informed alignment of the manual-grounded instructions and detailed reasoning examples with classifications, and a separate 236-utterance episode-disjoint pilot provided a replication check. The prompt was then frozen and applied to all 6,994 utterances without task-specific parameter fine-tuning or fitting to the full-corpus results.}

\begin{table}[htbp]
\centering
\small{\color{black}
\setlength{\tabcolsep}{4pt}
\begin{tabular}{lrrrr}
\toprule
\textbf{Model/ensemble} & \textbf{Accuracy} & \textbf{Macro-F1} & \textbf{Cohen's $\kappa$} & \textbf{MCC} \\
\midrule
BERT-base \citep{suresh2022talkmoves} & --- & .7312 & --- & .6716 \\
RoBERTa-base \citep{suresh2022talkmoves} & --- & .7158 & --- & .6605 \\
ELECTRA-base \citep{suresh2022talkmoves} & --- & .7128 & --- & .6447 \\
GPT-4, zero-shot \citep{ganesh2024prompting} & --- & .42 & --- & --- \\
\midrule
Claude Sonnet 5 & .7597 & .6780 & .6460 & .6497 \\
GPT-5.6 Terra & .7499 & .6740 & .6305 & .6321 \\
Gemini-3.8 Flash & .8067 & .7373 & .7124 & .7125 \\
Gemma4-31B & .7617 & .6795 & .6452 & .6460 \\
Qwen3.6-27B & .7395 & .6483 & .6095 & .6099 \\
\textbf{Our ensemble} & \textbf{.7754} & \textbf{.7002} & \textbf{.6661} & \textbf{.6669} \\
\bottomrule
\end{tabular}
}
\caption{\textcolor{black}{Five-category performance against the existing TalkMoves reference labels ($n=6{,}994$). Supervised student-model values are from \citet{suresh2022talkmoves}; the zero-shot GPT-4 macro-F1 is from \citet{ganesh2024prompting}, a naive in-context baseline with no fine-tuning; dashes indicate statistics not reported in the cited source. These baseline values are from prior work, were not evaluated in this study, and are included as documentation only.}}
\label{tab:talkmoves_performance}
\end{table}

\begin{table}[htbp]
\centering
{\color{black}
\begin{tabular}{lrr}
\toprule
\textbf{Agreement group} & \textbf{$n$} & \textbf{Accuracy} \\
\midrule
Unanimous five-model vote & 4,889 & 87.79\% \\
Split vote & 2,105 & 53.73\% \\
\bottomrule
\end{tabular}
}
\caption{\textcolor{black}{TalkMoves criterion accuracy by ensemble agreement status.}}
\label{tab:talkmoves_agreement}
\end{table}

{\color{black}Table \ref{tab:talkmoves_performance} reports accuracy=.7754, macro-F1=.7002, Cohen's $\kappa=.6661$, and MCC=.6669 for the ensemble, exceeding four of its five constituent models and approaching the supervised, fine-tuned BERT-base baseline (.7312 macro-F1) reported by \citet{suresh2022talkmoves}, while substantially exceeding the naive zero-shot GPT-4 baseline (.42 macro-F1) reported for the same task by \citet{ganesh2024prompting}. The agreement table shows that unanimous items were more accurate than split-vote items (87.79\% versus 53.73\%). Thus, agreement is empirically useful for review prioritization on this external-reference task, but it remains distinct from correctness and does not directly establish the accuracy of labels in the classroom corpus analyzed here.}

\subsubsection{Human-LLM Classification Comparison}

To validate the reliability of our automated classification framework, we additionally applied a human annotation assessment and examined the consistency between human judgments and LLM-generated results. Specifically, we created standardized labeling guidelines that align precisely with the classification criteria used in our LLM prompts, ensuring methodological consistency between the automated and manual evaluation processes.


Eight human annotators participated in the validation study, including seven graduate students and one senior undergraduate student, all majoring in statistics and data science at a large research university and possessing formal training in quantitative methods. All annotators completed structured training and practice sessions prior to the study to ensure a shared understanding of the coding criteria and procedures. The annotation process was conducted through a standardized web-based platform that was specifically developed to promote consistency in labeling, reduce potential sources of inter-annotator variability, and maintain uniform working conditions across participants. This design was intended to ensure that differences in annotations would reflect substantive judgment rather than inconsistencies in instructions, interface, or experience.

The human annotation dataset comprised five strategically selected dialogue episodes that maximized the representation of the participants in classroom interactions. This focused subset was selected due to time and resource constraints, as full annotation of the complete dataset collected during the five-month instructional period would have exceeded the available capacity. To establish quantitative measures of annotation reliability, we implemented both Cohen's $\kappa$ and Fleiss' $\kappa$ coefficients for systematic assessment of inter-annotator agreement patterns. The heatmap visualization displaying pairwise Cohen's $\kappa$ values between human annotators and LLMs is presented in Section \ref{supp-sec:uncertainty} of the Supplementary Material. \textcolor{black}{Beyond utterance-level agreement, we also compared the constructed networks directly: the EXP network built from the final ensemble agrees with the eight-annotator human-consensus EXP network at edge F1 $=.913$, with the mean human-to-ensemble edge F1 (.913) falling just above the mean human-to-human edge F1 (.877). The full network-level comparison, including edge-weight and centrality correlations, is reported in Section \ref{supp-sec:uncertainty} of the Supplementary Material.}

\subsubsection{Uncertainty Quantification}

\paragraph*{Inter-Model Uncertainty Quantification} To systematically quantify the degree of disagreement among models and \textcolor{black}{identify utterances that can be prioritized for review}, we implemented Shannon entropy \citep{shannon} as our primary uncertainty measure. This information-theoretic approach enables consistent measurement of classification uncertainty by analyzing the probability distribution of label choices across multiple LLM outputs, where each label's probability is calculated as the proportion of models selecting that label. For a given utterance with classification probabilities $\{p_1, p_2, \dots, p_n\}$ distributed across $n$ possible classification labels, the Shannon entropy, $H(X)$, is defined as:
\begin{equation} \label{eq:shannon}
    H(X) = -\sum_{i=1}^{n} p_i \log_2 p_i, 
\end{equation}
where the probability $p_i$ for each label category is empirically estimated through ensemble voting as:
\begin{equation}
    p_i = \frac{\text{number of models selecting label } i}{\text{total number of models}}. 
\end{equation}
The resulting entropy values operate within a defined range from 0, indicating perfect consensus among all models, to $\log_2(n)$, representing maximum uncertainty where models are equally divided across all possible labels. Utterances exhibiting high entropy values signal substantial inter-model disagreement and are systematically flagged as priority candidates for expert human annotation. This entropy-driven selection strategy enables efficient allocation of limited annotation resources while maintaining rigorous quality standards, effectively reducing the computational and human burden associated with comprehensive manual review processes.

\begin{figure}[!ht]
    \centering
    \includegraphics[width=.9\textwidth]{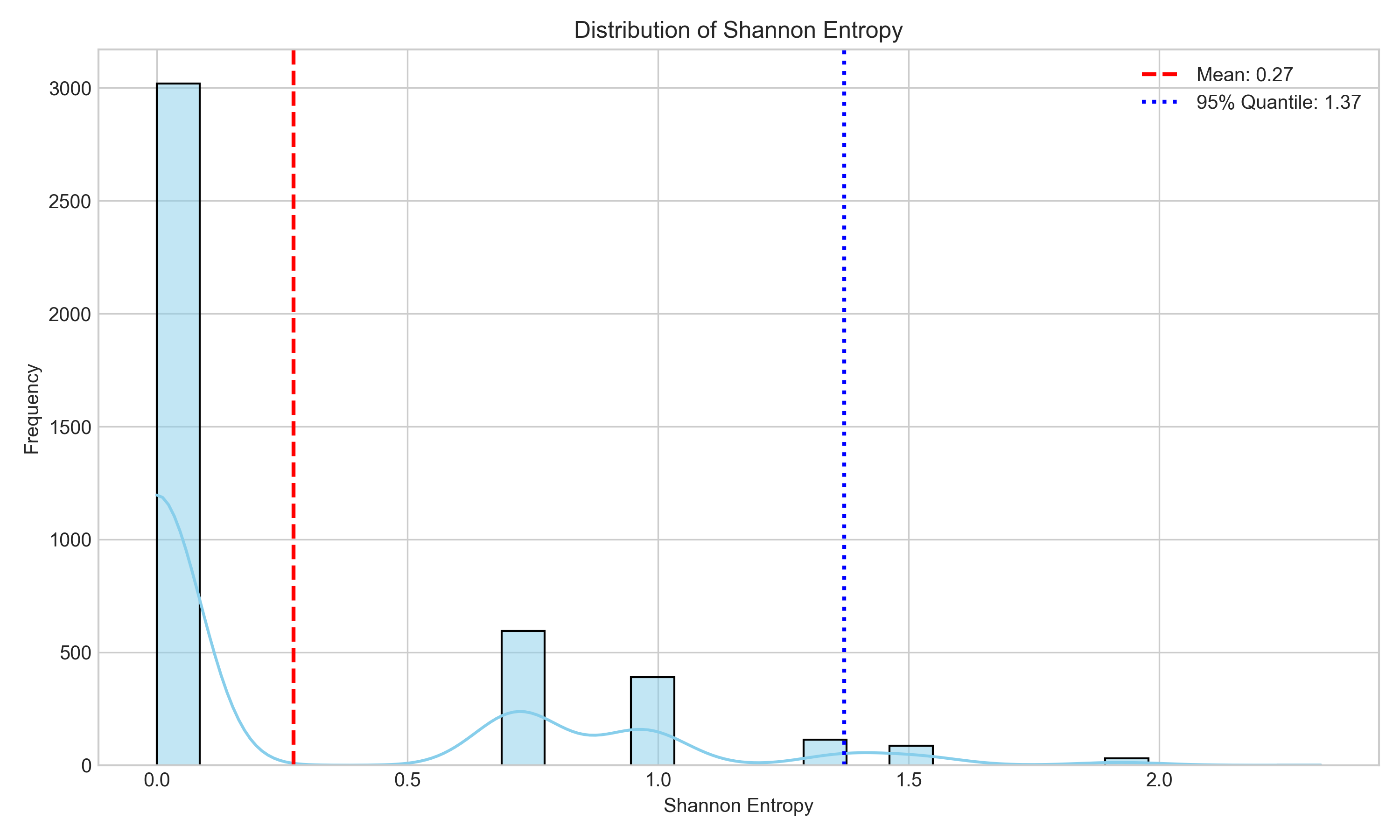}
    \caption{Distribution of Shannon entropy across \textcolor{black}{five} LLMs for inter-model uncertainty quantification}
    \label{fig:entropy_llms}
\end{figure}

{\color{black}Figure \ref{fig:entropy_llms} illustrates the distribution of Shannon entropy values derived from classification results across our five-model ensemble. The analysis of 4,235 total utterances reveals that 71.29\% (3,019 utterances) achieved complete consensus among all models, indicating substantial inter-model agreement on this task. The calculated 95th percentile entropy threshold of 1.370951 represents classification scenarios where three of the five models converge on identical label assignments, indicating that the vast majority of utterances generate strong consensus patterns across the ensemble. Conversely, the top 5\% of utterances exhibiting the highest entropy values reveal significant inter-model disagreement, providing a principled basis for prioritizing cases for review.}

\paragraph*{Intra-Model Uncertainty Quantification} 
To evaluate model confidence with greater precision, we implemented a complementary intra-model uncertainty assessment comprising sequence-level and token-level probability analyses. These two approaches target distinct aspects of model confidence. Sequence-level analysis evaluates the probability assigned to the complete classification label as a coherent linguistic unit, capturing overall labeling certainty. Token-level analysis, by contrast, examines the probability distribution at each generated token, revealing where in the generation process the model is most uncertain about its classification decision. We applied Shannon entropy as our primary uncertainty quantification metric for both analytical approaches. Importantly, these two dimensions of uncertainty are complementary rather than redundant: an utterance may exhibit high sequence-level uncertainty because the complete label string receives low overall probability, yet show low token-level entropy if the model's decision at each individual token is relatively unambiguous.

{\color{black}Sequence-level probability diagnostics were available for the two locally served open-weight models. Using $P_{\max}<.99$ as a descriptive review flag, 73 of 3,393 utterances (2.15\%) were flagged for Gemma4-31B and 802 of 3,393 (23.64\%) for Qwen3.6-27B. These model-specific values are not generalized to the three commercial models or to the full ensemble.}

\begin{figure}[!ht]
    \centering
    \includegraphics[width=.9\textwidth]{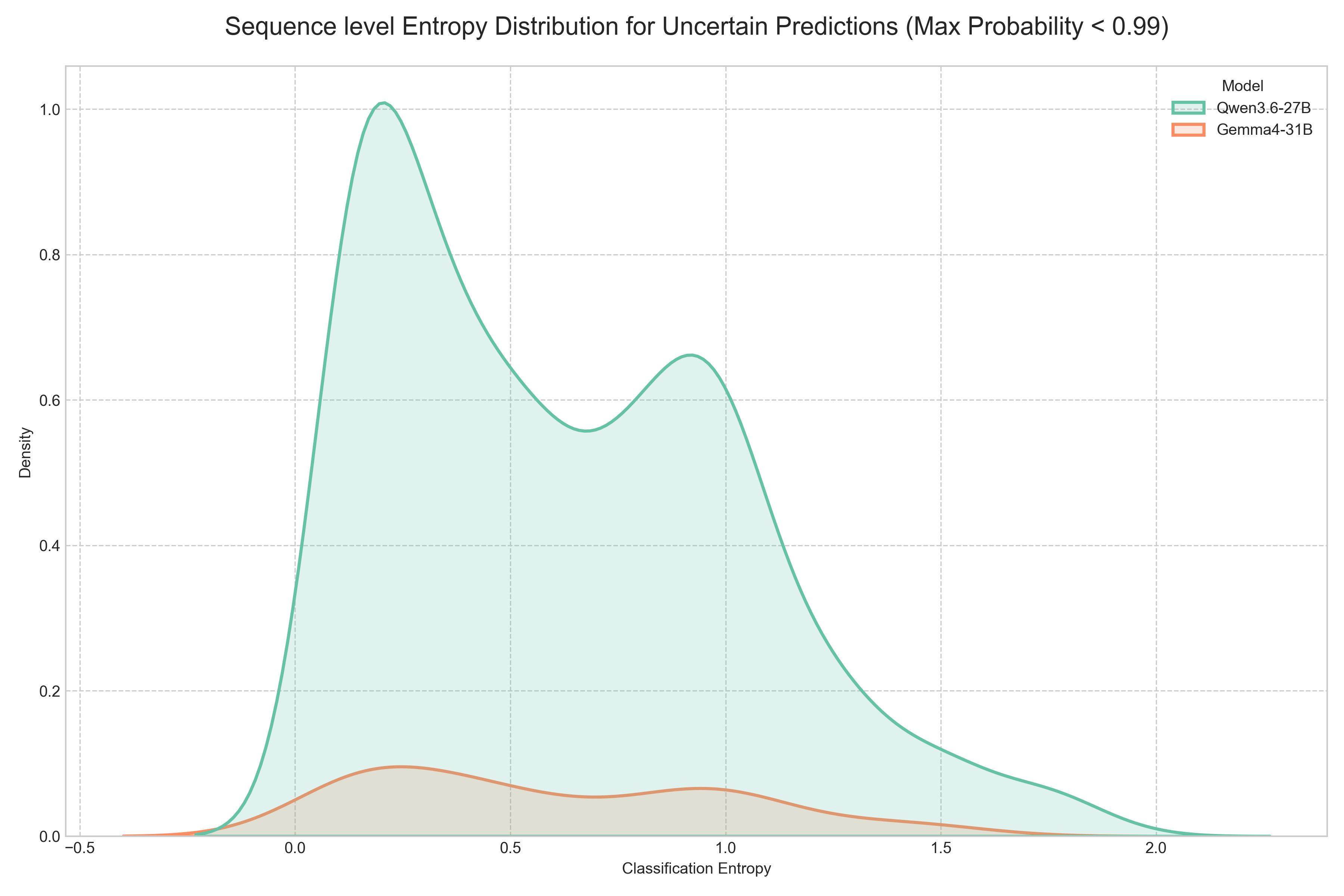}
    \caption{
    \textcolor{black}{Sequence-level Shannon entropy for the two locally served open-weight models, restricted to utterances with maximum normalized sequence probability below .99.}
    }
    \label{fig:sequence_entropy}
\end{figure}

At the token level, we examined the model's decision-making process through a two-stage hierarchical analysis, first evaluating primary category selection (Explain vs.\ Engage) and then analyzing specific engagement level assignment (Low, Medium, and High). The token-level results revealed that even within these difficult cases, models generally identified primary categories with high confidence (lower entropy) but exhibited substantially greater uncertainty (broader and higher entropy distributions) when determining specific engagement levels. This pattern is consistent with the inherent difficulty of fine-grained classification tasks and underscores the value of decomposing uncertainty by classification stage. Detailed technical specifications, mathematical formulations, and empirical examples illustrating these patterns are available in Section \ref{supp-sec:uncertainty} of the Supplementary Material.

\begin{figure}[!ht]
    \centering
    \includegraphics[width=0.8\textwidth]{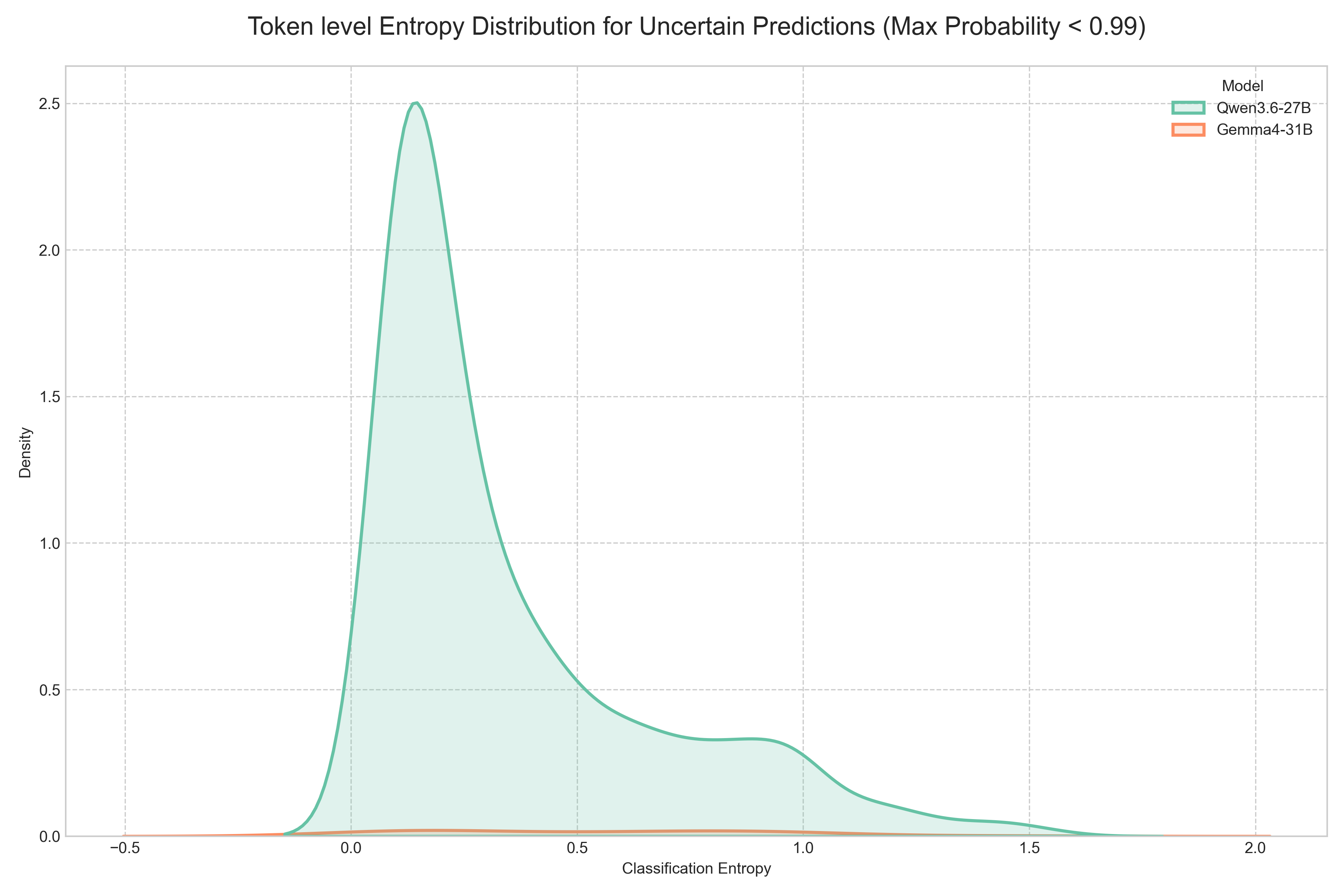}
    \caption{
    \textcolor{black}{Token-level Shannon entropy for primary-category selection across Gemma4-31B and Qwen3.6-27B, restricted to locally available probability diagnostics with maximum probability below .99.}
    }
    \label{fig:token_entropy_comparison}
\end{figure}

\begin{figure}[!ht]
    \centering
    \includegraphics[width=0.8\textwidth]{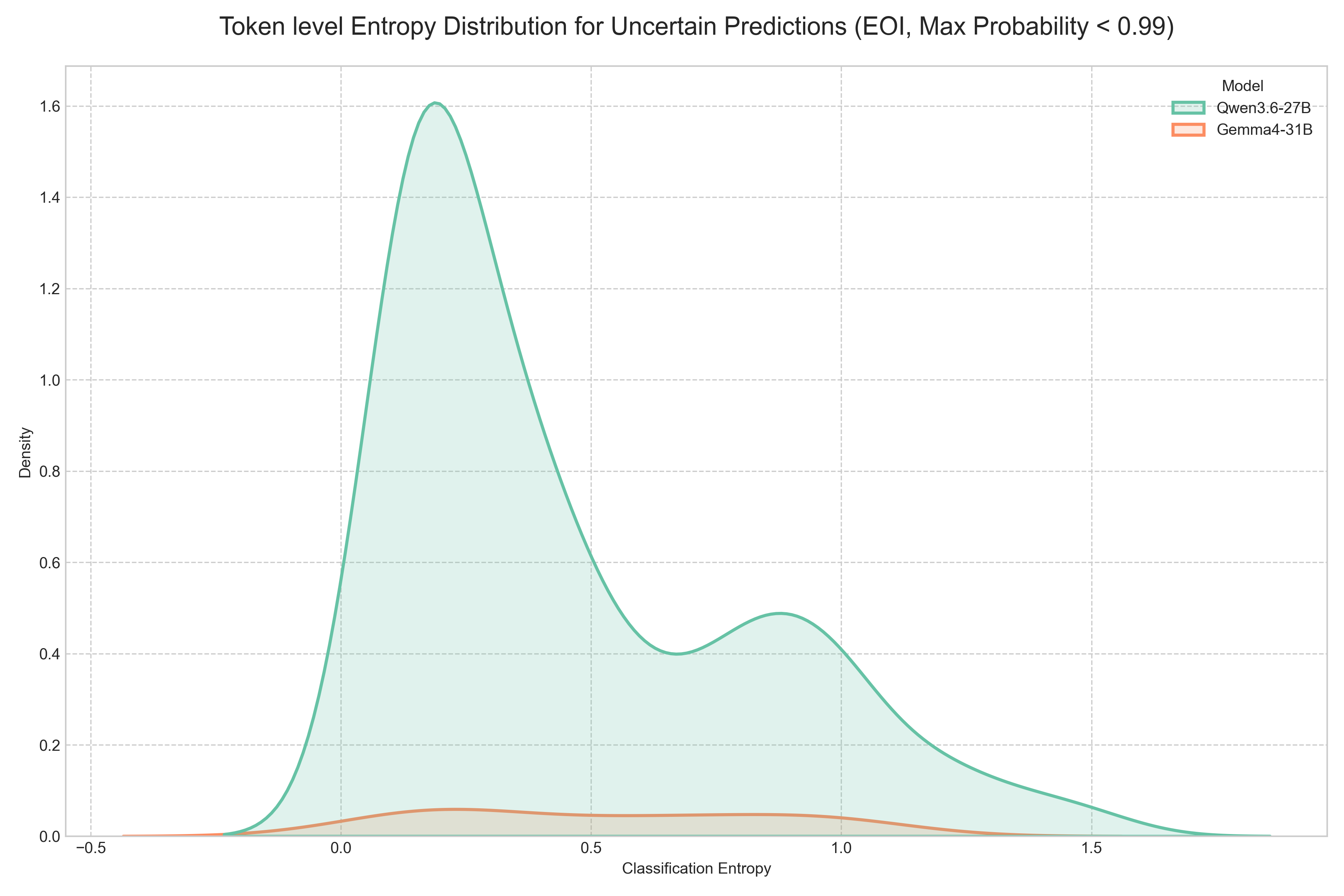}
    \caption{
    \textcolor{black}{Token-level Shannon entropy for engagement-level selection across Gemma4-31B and Qwen3.6-27B, conditional on utterances classified as Engage Others' Idea and restricted to locally available probability diagnostics with maximum probability below .99.}
    }
    \label{fig:token_engage_entropy_comparison}
\end{figure}

\subsubsection{Summary}

The validation framework supports the reliability of LLM-based classification through complementary consistency, \textcolor{black}{external criterion,} and uncertainty metrics. Ensemble analysis across the full dataset demonstrated substantial agreement (\textcolor{black}{Fleiss' $\kappa=.7164$, with 71.29\% unanimous agreement}), \textcolor{black}{and comparative evaluation on TalkMoves, an external corpus with existing reference labels, showed accuracy=.7754 and macro-F1=.7002, with lower accuracy for split than unanimous votes}. It should be noted that this human comparison is based on a limited subset of the data, and generalizing these agreement levels to the full dataset warrants caution. Uncertainty was evaluated at both inter-model and intra-model levels \textcolor{black}{as a review-prioritization diagnostic rather than as evidence of accuracy, while flagging} items for expert review. In particular, token-level analysis revealed that uncertainty arises primarily during fine-grained engagement level selection rather than primary category identification, providing actionable guidance for targeted human review. Together, these results indicate that combining ensemble consensus, \textcolor{black}{external criterion performance,} and internal confidence metrics \textcolor{black}{provides complementary evidence for constructing interaction networks, though none alone establishes correctness on the full Webb corpus or unrestricted generalizability}.

Although \textcolor{black}{examined} here using classroom dialogue, this methodology \textcolor{black}{may be adapted to other conversational settings only with domain-specific coding definitions, prompt alignment, network-construction rules, and validation. The present evidence supports its use in the educational-discourse settings examined here rather than unrestricted transfer to clinical, organizational, or other domains.}

\section{Network Construction and Illustrative Applications}
\label{sec:applications}

Network analysis provides a natural framework for representing the relational structures inherent in conversational data \citep{grunspan2014understanding, bokhove2018exploring}. By encoding classified interactions as directed edges between participants, researchers can construct social networks amenable to a wide range of analytical approaches, from descriptive characterization through centrality metrics \citep{williams2019linking} to statistical modeling of latent social structures \citep{lsm, hoff2005bilinear} and mediation analysis \citep{liu2021social, networkmed}. Building upon the utterance classification framework \textcolor{black}{developed and evaluated} in Section \ref{sec:prompt}, this section first describes how to construct structured networks from the classified conversational data, and then illustrates their analytical potential through two complementary applications: network centrality analysis for describing participant roles, and network mediation analysis for statistically examining how network structures mediate relationships between individual attributes and outcomes. These applications are presented as illustrative cases using classroom interaction data; \textcolor{black}{adaptation of the network-construction procedure to other conversational settings requires domain-specific coding and construction rules.}

The classroom dataset used throughout this section comprises demographic and performance records for the 20 third-grade students (9 male, 11 female) introduced in Section \ref{sec:prompt}. To track educational growth alongside interaction patterns, the original study assessed mathematics performance before and after the recorded lessons. The pre-test, based on the California Standards Test (CST) \citep{starprogram}, yielded a mean score of 369.35 (SD = 54.31). Due to a statewide transition in assessment frameworks during the study period, the post-test utilized an author-developed measure designed to assess multiplication and division problem-solving strategies \citep{webb2023learning}, yielding a mean score of 15.08 out of 24 (SD = 6.49).

\subsection{Network Construction from Classified Utterances}
After classifying utterances using prompt engineering, we created a directed network representation to systematically analyze student interaction patterns. The representation of student interaction patterns as networks enables both visualization and systematic analysis of classroom dynamics.

We formalized student interactions using an adjacency matrix, where the matrix elements indicate the frequency of specific interaction types between two students during the observation period. We then visualized the constructed network structure using nodes to represent individual students and directed weighted edges between nodes to represent the nature and frequency of the interactions between the two students. For explanatory (EXP) utterances, directed edges are constructed from the speaker to all other members of the same small group, reflecting the broadcast nature of explanations within group discussions. For engagement (EOI) utterances, the LLM identifies the specific student whose idea is being engaged with based on the conversational context, and a directed edge is formed from the engaging student to that target student.

{\color{black}To capture the varying intensity of student engagement more precisely, we implemented a differential weighting scheme for EOI interactions. This ordinal weighting reflects the hierarchical nature of the engagement levels defined in the coding scheme, approximating the inverse of the relative frequency of each level in Table 1 (124 high-, 187 medium-, and 359 low-engagement utterances) so that the less frequent, higher-engagement interactions receive proportionally more weight: medium-level engagements received double weight, and high-level engagements received triple weight relative to low-level engagements, while explanatory interactions maintained equal weights.}

While this network construction process allows us to represent valuable aspects of classroom interactions, analyzing classroom conversation data in a network also presents unique analytical challenges. Unlike conventional social network datasets, our classroom interactions are often characterized by limited and sparse connections, as conversations frequently occur within small groups during specific question discussions. To overcome the sparsity of individual lesson networks and facilitate a more comprehensive analysis of overall interaction patterns, we aggregated data across all six lessons to construct a single composite network. {\color{black}Unlike conventional social networks in which a tie is observed directly, each edge here is the product of a coded utterance, a construction rule and a reference-target judgment. Network validity is therefore inherited from classification validity, which is why the classification stage is evaluated against external and human criteria in Section 2 before network results are interpreted.}

\subsubsection{Network Graphs}

\begin{figure}[htbp]
    \centering
    \begin{subfigure}[b]{0.46\textwidth}
        \centering
        \includegraphics[width=\textwidth]{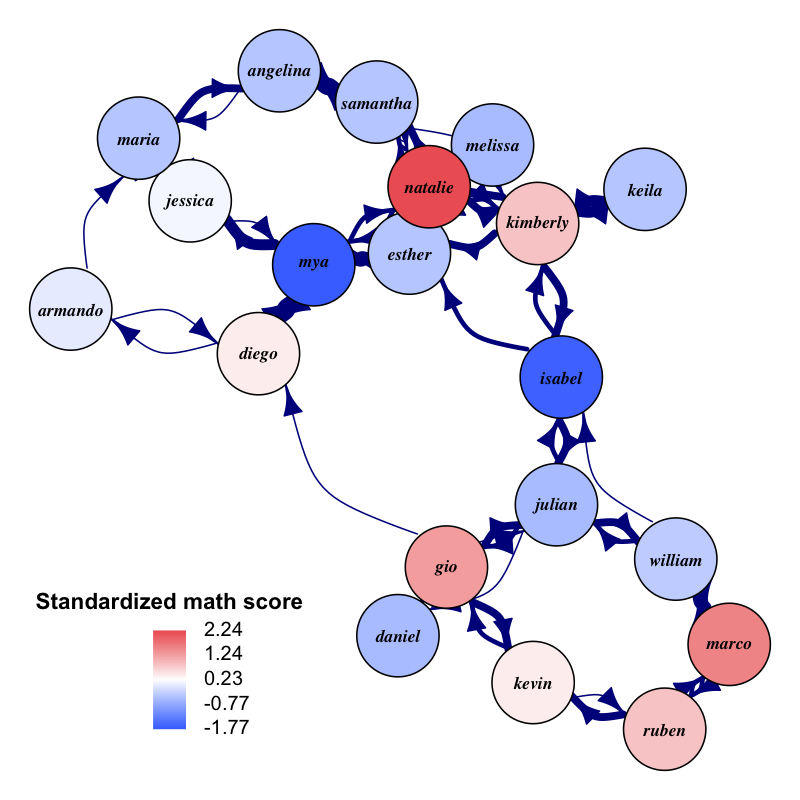}
        \caption{Explain Own Idea Network, Pre-test}
        \label{fig:exp_pre}
    \end{subfigure}
    \hspace{0.05\textwidth}
    \begin{subfigure}[b]{0.46\textwidth}
        \centering
        \includegraphics[width=\textwidth]{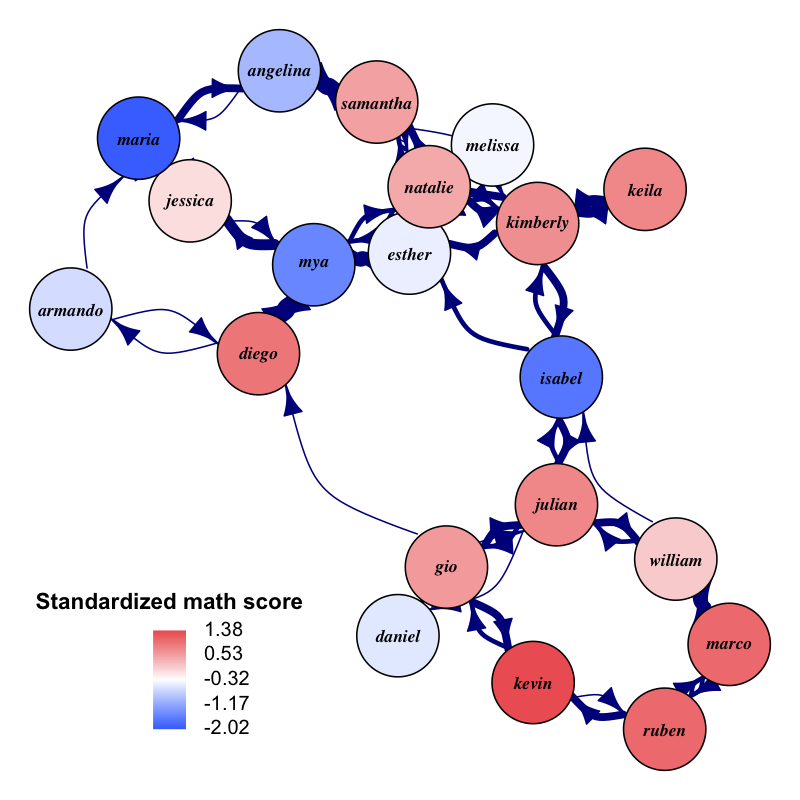}
        \caption{Explain Own Idea Network, Post-test}
        \label{fig:exp_post}
    \end{subfigure}
    
    \vspace{\baselineskip}

    \begin{subfigure}[b]{0.46\textwidth}
        \centering
        \includegraphics[width=\textwidth]{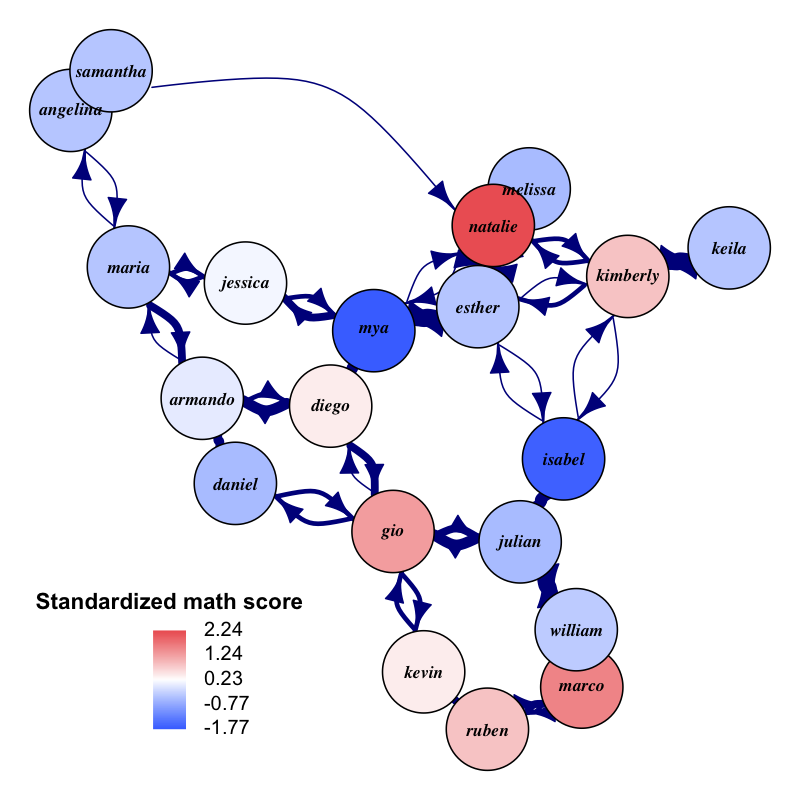}
        \caption{Engage Others' Idea Network, Pre-test}
        \label{fig:eoi_pre}
    \end{subfigure}
    \hspace{0.05\textwidth}
    \begin{subfigure}[b]{0.46\textwidth}
        \centering
        \includegraphics[width=\textwidth]{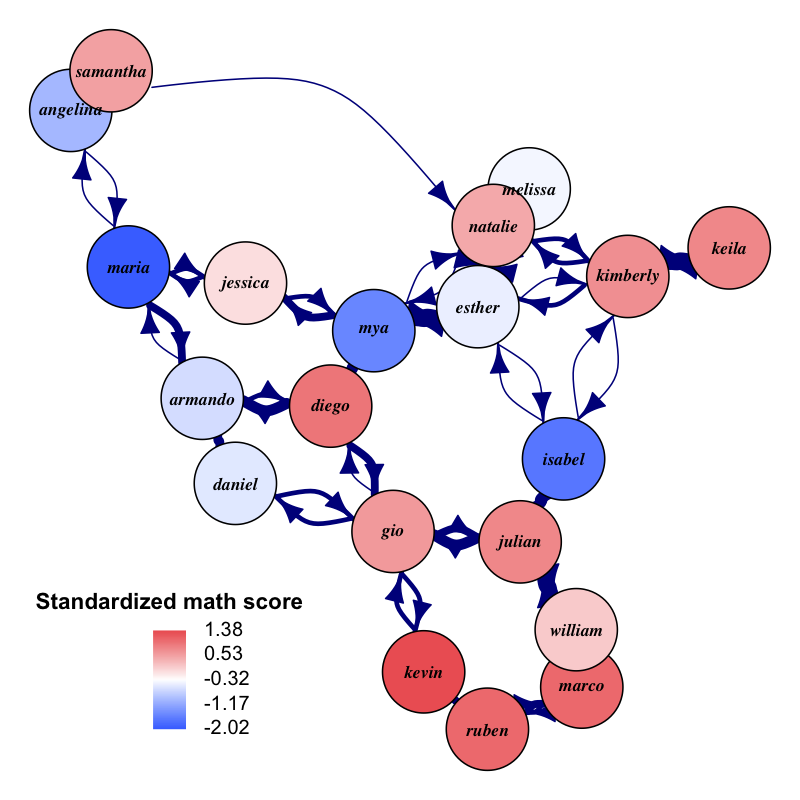}
        \caption{Engage Others' Idea Network, Post-test} 
        \label{fig:eoi_post}
    \end{subfigure}
    \caption{Explain Own Idea and Engage Others' Idea Networks. Edge weights in each network represent the frequency of corresponding utterances, and node colors represent math test scores  (red = higher, blue = lower). The left-hand figures show pre-test results, and the right-hand figures show post-test results. The networks are constructed using data aggregated from six lessons.} 
    \label{fig:all_networks} 
\end{figure}

Figure \ref{fig:all_networks} illustrates these network transformations, where each node represents a student, and the edges represent their interactions. The node colors indicate students' standardized mathematics performance, utilizing a spectrum from blue (lower scores) to red (higher scores). The edge thickness corresponds to the interaction frequency between pairs of students.

{\color{black}Figure \ref{fig:all_networks} visualizes the EXP and EOI networks reconstructed from the corrected five-model ensemble. The panels show substantial heterogeneity in both interaction frequency and direction across students, alongside changes in test-score coloring from pre-test to post-test. Because these graphs are descriptive, they are used to motivate the centrality and network-mediation analyses rather than to attribute score changes to particular interaction partners.}

\subsection{Descriptive Analysis via Network Centrality}

We examined network centrality, using metrics designed for directed weighted networks, accounting for both directional flow and interaction intensity. To comprehensively capture student engagement, influence, and brokerage roles, we applied multiple measures: PageRank \citep{pagerank}, hub-authority scores \citep{hubauth}, \textcolor{black}{in-strength and out-strength}, betweenness \citep{freeman1977set}, closeness \citep{bavelas1948mathematical}, and eigenvector centrality \citep{bonacich1972factoring}. Each centrality measure provides a distinct perspective on the structural positions students occupy within the classroom interaction network.

\begin{table}[!ht]
\centering
\small{\color{black}
\setlength{\tabcolsep}{3pt}
\begin{tabular}{lrrrrrrrr}
\toprule
    & \textbf{PageRank} & \textbf{In-strength} & \textbf{Out-strength} & \textbf{Betweenness} & \textbf{Closeness} & \textbf{Eigen} & \textbf{Hub} & \textbf{Authority}\\
\midrule
Angelina & \textbf{0.0857} & 22 & 30 & 8 & 0.1496 & 0.0967 & 0.2095 & 0.0697\\
Armando & 0.0149 & 4 & 7 & 0 & 0.1030 & 0.0457 & 0.0420 & 0.0185\\
Daniel & 0.0202 & 8 & 5 & 0 & 0.1073 & 0.0005 & 0.0029 & 0.0109\\
Diego & 0.0437 & 22 & 20 & 61 & 0.1802 & 0.4026 & 0.1541 & \textbf{0.4311}\\
Esther & 0.0819 & \textbf{53} & 34 & 41 & \textbf{0.1943} & \textbf{0.8322} & \textbf{0.3132} & \textbf{1.0000}\\
Gio & 0.0434 & 13 & 26 & 74 & 0.1583 & 0.0010 & 0.0548 & 0.0029\\
Isabel & 0.0383 & 18 & 20 & \textbf{132} & 0.1885 & 0.0204 & 0.1416 & 0.0911\\
Jessica & 0.0631 & 30 & 14 & 41 & 0.1706 & \textbf{0.5160} & 0.0450 & \textbf{0.5618}\\
Julian & 0.0590 & 26 & 24 & 129 & 0.1705 & 0.0064 & 0.0171 & 0.0584\\
Keila & 0.0181 & 11 & 18 & 0 & 0.1634 & 0.0260 & 0.0447 & 0.1185\\
Kevin & 0.0321 & 16 & 9 & 13 & 0.1219 & 0.0002 & 0.0004 & 0.0138\\
Kimberly & 0.0486 & 32 & \textbf{43} & \textbf{144} & \textbf{0.1953} & 0.0833 & \textbf{0.3595} & 0.1073\\
Marco & 0.0344 & 12 & 23 & 22 & 0.1236 & 0.0004 & 0.0026 & 0.0032\\
Maria & 0.0561 & 15 & 23 & 16 & 0.1493 & 0.1558 & 0.1965 & 0.0298\\
Melissa & 0.0204 & 10 & 17 & 0 & 0.1438 & 0.0987 & 0.2615 & 0.0858\\
Mya & \textbf{0.1212} & \textbf{56} & \textbf{62} & \textbf{139} & \textbf{0.1956} & \textbf{1.0000} & \textbf{1.0000} & 0.4117\\
Natalie & 0.0497 & 28 & \textbf{35} & 90 & 0.1600 & 0.2940 & 0.2832 & 0.3672\\
Ruben & 0.0273 & 10 & 14 & 3 & 0.1041 & 0.0001 & 0.0030 & 0.0005\\
Samantha & \textbf{0.0916} & \textbf{43} & 17 & 24 & 0.1560 & 0.1542 & 0.0429 & 0.3214\\
William & 0.0500 & 27 & 15 & 51 & 0.1377 & 0.0020 & 0.0149 & 0.0064\\
\bottomrule
\end{tabular}
}
\caption{Centrality measures for the EXP (Explain Own Idea) network. The top three values in each column are highlighted in bold.}
\label{tab:expcent}
\end{table}

\begin{table}[!ht]
\centering
\small{\color{black}
\setlength{\tabcolsep}{3pt}
\begin{tabular}{lrrrrrrrr}
\toprule
    & \textbf{PageRank} & \textbf{In-strength} & \textbf{Out-strength} & \textbf{Betweenness} & \textbf{Closeness} & \textbf{Eigen} & \textbf{Hub} & \textbf{Authority}\\
\midrule
Angelina & 0.0422 & 55 & 36 & 34 & 0.1501 & 0.0237 & 0.0002 & 0.0128\\
Armando & 0.0615 & 65 & 27 & 24 & 0.3809 & 0.7483 & 0.0646 & 0.0203\\
Daniel & 0.0556 & 28 & 28 & 16 & 0.2968 & 0.2983 & 0.0048 & 0.0276\\
Diego & 0.0457 & 61 & \textbf{84} & 95 & \textbf{0.4306} & \textbf{1.0000} & 0.0124 & \textbf{0.9155}\\
Esther & 0.0567 & \textbf{80} & 30 & \textbf{121} & 0.3804 & \textbf{0.9821} & 0.0071 & \textbf{1.0000}\\
Gio & \textbf{0.0733} & \textbf{78} & 47 & 96 & \textbf{0.3943} & 0.2390 & 0.0389 & 0.0152\\
Isabel & 0.0527 & 51 & 22 & 70 & 0.3121 & 0.0000 & 0.0527 & 0.0241\\
Jessica & 0.0235 & 31 & 27 & 88 & 0.3170 & 0.3412 & 0.0017 & \textbf{0.3296}\\
Julian & \textbf{0.0902} & 68 & \textbf{93} & \textbf{109} & 0.3474 & 0.0000 & 0.0155 & 0.0412\\
Keila & 0.0427 & 36 & 17 & 0 & 0.2175 & 0.1516 & 0.0014 & 0.0861\\
Kevin & 0.0363 & 34 & 35 & 17 & 0.2669 & 0.0000 & 0.0020 & 0.0080\\
Kimberly & \textbf{0.0691} & 39 & 60 & 37 & 0.2440 & 0.2239 & \textbf{0.1347} & 0.0086\\
Marco & 0.0536 & \textbf{69} & 54 & 19 & 0.2751 & 0.0000 & 0.0032 & 0.0188\\
Maria & 0.0260 & 20 & 33 & 66 & 0.2887 & 0.1138 & 0.0435 & 0.0027\\
Melissa & 0.0313 & 26 & 41 & 1 & 0.2571 & 0.1787 & \textbf{0.1642} & 0.0096\\
Mya & 0.0610 & 63 & \textbf{128} & \textbf{189} & \textbf{0.4098} & \textbf{0.9256} & \textbf{1.0000} & 0.0108\\
Natalie & 0.0436 & 50 & 45 & 20 & 0.2296 & 0.2849 & 0.0225 & 0.2185\\
Ruben & 0.0371 & 34 & 45 & 4 & 0.2383 & 0.0000 & 0.0053 & 0.0015\\
Samantha & 0.0384 & 31 & 52 & 0 & 0.1424 & 0.0138 & 0.0101 & 0.0001\\
William & 0.0595 & 62 & 77 & 47 & 0.2981 & 0.0000 & 0.0192 & 0.0077\\
\bottomrule
\end{tabular}
}
\caption{Centrality measures for the EOI (Engage Others' Idea) network. The top three values in each column are highlighted in bold.}
\label{tab:eoicent}
\end{table}

Tables \ref{tab:expcent} and \ref{tab:eoicent} demonstrate how multiple centrality measures provide complementary insights into different patterns of student engagement within classroom interaction networks. 
{\color{black}The corrected networks indicate distinct structural roles across interaction types. In EXP, Mya has the highest PageRank (.1212), in-strength (56), out-strength (62), eigenvector centrality (1.0000), and hub score (1.0000), while Esther has the highest authority score (1.0000). Kimberly, Mya, and Isabel have the three largest betweenness values. In EOI, Julian has the highest PageRank (.0902), Diego has the highest eigenvector centrality (1.0000), Mya has the highest out-strength (128), betweenness (189), and hub score (1.0000), and Esther again has the highest in-strength (80) and authority score (1.0000). These differences show that prominence depends on both the interaction type and the centrality definition.}

These exploratory observations also suggest that high network centrality does not necessarily correspond to improved mathematical performance, indicating that prominent positioning within classroom interaction networks alone may be insufficient to predict learning outcomes. Mya's case exemplifies this pattern: despite maintaining the highest centrality across multiple measures in both networks, her mathematics scores remained below the class average throughout the study period. This finding suggests that the relationship between network position and academic success may operate through more nuanced mechanisms than global connectivity metrics capture. For instance, the academic caliber of immediate interaction partners or the quality of local network structures may exert greater influence on learning outcomes than overall network prominence. It should be noted that these centrality-based interpretations are strictly exploratory and do not support causal inferences; a more formal statistical investigation of these relationships is undertaken in the following section.

\subsection{Statistical Modeling via Network Mediation}

While the centrality analysis in the preceding section identified structurally prominent students, centrality metrics alone do not reveal the mechanisms through which network positions are associated with learning outcomes. To move beyond description toward a more formal investigation of these relationships, we employ network mediation analysis \citep{liu2021social, networkmed, di2022networks}. This framework treats the network structure as a mediating variable, enabling decomposition of the total association between an individual attribute and an outcome into direct and indirect (network-mediated) components. A key challenge in implementing this approach is representing the high-dimensional network structure in a form suitable for use as a mediator. We address this by adopting a latent space model (LSM) approach \citep{lsm, hoff2005bilinear, sewell2016latent}, which represents each student's position within the network as a low-dimensional latent vector, capturing underlying social dynamics wherein inter-student distances reflect the probability and intensity of their interactions.

\subsubsection{Conceptual Framework: Network as Mediator} 

Our study examines how classroom interaction networks mediate the influence of gender on mathematics achievement. Prior research suggests that gender differences in mathematics performance are largely shaped by social and contextual factors rather than biological ones \citep{lindberg2010new, else2010cross, ghasemi2019gender}. Classroom interaction networks represent a concrete manifestation of these social factors: the patterns of who explains ideas to whom and who engages with whose reasoning create differential learning opportunities that may vary systematically by gender. This perspective motivates treating the network structure as a potential mediator of the gender--performance relationship.

Based on this perspective, we hypothesize that the gender gap in mathematics performance ($X$ on $Y$) can be differentiated based on how individual students interact with others during mathematics discussion classes. Specifically, we apply network mediation analysis \citep{liu2021social, networkmed, di2022networks}, using students' interaction structures as mediators ($M$) \textcolor{black}{of the gender ($X$)--mathematics performance ($Y$) relationship}. By incorporating estimated latent positions ($z_i$, $w_j$) as mediators, our goal is to investigate the indirect effects of gender on mathematics performance through the network structure.

To address potential confounding effects, we incorporated students' prior mathematical proficiency as a confounding variable in our mediation analysis. We use California Standards Test (CST) performance, with students scoring 350 or above the state-designated proficiency threshold classified as high-performing and those below this threshold categorized as requiring additional support. This approach addresses non-linear relationships observed between pre-test and post-test performance measures while systematically controlling for baseline achievement differences that could otherwise confound the relationship between gender, network position, and learning outcomes. Detailed justification for our choice of the confounding variable is provided in Section \ref{supp-sec:confounding} of the Supplementary Material.

\begin{figure}
    \centering
    \includegraphics[width=0.7\linewidth]{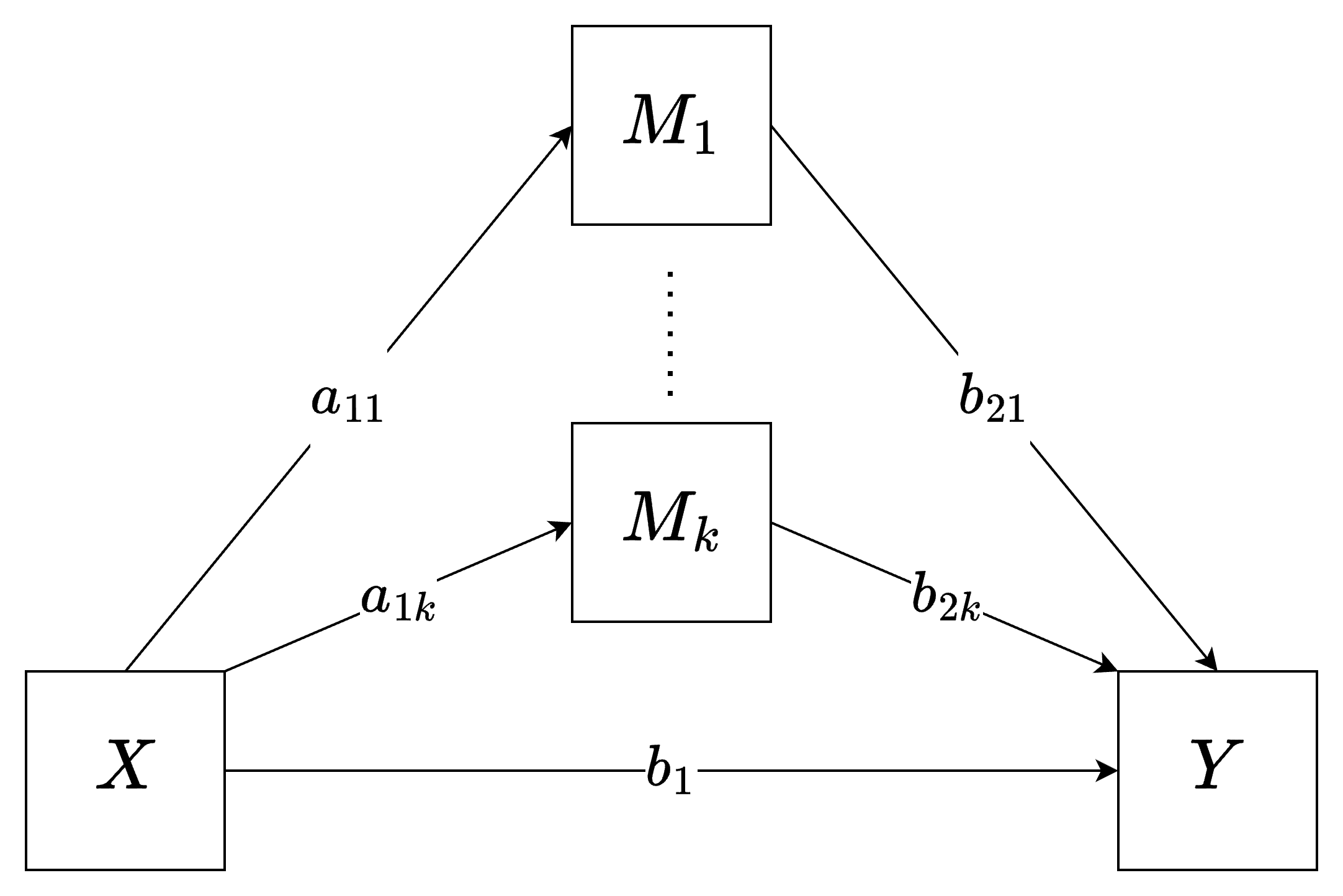}
    \caption{Multiple mediator model for network mediation analysis}
    \label{fig:networkmed}
\end{figure}

Figure \ref{fig:networkmed} presents a diagram of a multiple mediator model for network mediation analysis. In this model, $M_1$ through $M_k$ represent the latent positions of network nodes embedded in a latent space. The paths $a_{11}$ through $a_{1k}$ denote the effects of $X$ on the mediators, while the paths $b_{21}$ through $b_{2k}$ capture the effects of the mediators on $Y$. The direct effect of $X$ on $Y$ is represented by path $b_1$. By linking these latent positions to independent and dependent variables through regression frameworks, we can decompose total effects into direct and indirect components mediated by the underlying network structure.

\subsubsection{Stage 1: Latent Position Estimation via AMEN Model}

To obtain the latent positions that serve as mediators, we employ the AMEN model \citep{amen}. 
The AMEN model extends the latent space model \citep[LSM;][]{lsm} to directed networks by assigning each node distinct latent positions as a sender and as a receiver. This asymmetric representation is well suited to classroom interaction data, where the direction of each utterance---who speaks to whom---carries substantive meaning that a symmetric model would obscure.

A further consideration is that our network data consist of interaction counts that exhibit severe overdispersion: the variance of edge weights substantially exceeds their mean (see Section \ref{supp-sec:overdispersion} of the Supplementary Material for a detailed assessment). Standard Poisson regression, which constrains the variance to equal the mean, cannot adequately capture this excess variability. We therefore adopt a negative binomial regression structure for the AMEN model, which introduces a dispersion parameter that allows the conditional variance to vary independently of the mean. The resulting model can be written as:

\begin{equation}\label{amen}
    \log(\mu_{ij}) = \alpha_i + \beta_j + z_i^Tw_j,
\end{equation}
\noindent 

where $y_{ij}$ denotes the observed interaction count from node $i$ (sender) to node $j$ (receiver), modeled as $y_{ij} \sim \mbox{NB}(\mu_{ij}, r)$. Here, $\mu_{ij}$ is the conditional mean and $r$ is the dispersion parameter, with smaller values of $r$ corresponding to greater overdispersion. The additive components $\alpha_i$ and $\beta_j$ are sender and receiver random effects, respectively, capturing node-specific tendencies to initiate or attract interactions. The vectors $z_i$ and $w_j$ are $d$-dimensional latent feature vectors for node $i$ as a sender and node $j$ as a receiver. The multiplicative term $z_i^T w_j$ captures dyad-specific affinity beyond what the additive effects explain, enabling the model to represent complex higher-order network dependencies such as transitivity, clustering, and degree heterogeneity.

\subsubsection{Stage 2: Network Mediation Model Specification}

We also consider the effects of the predictor-mediator interaction to account for the potential heterogeneity in mediator effects at different predictor levels \citep{vanderweele2009conceptual, valeri2013mediation}. In this case, indirect effects, mediated through the latent network structure, vary by gender ($x$). By modeling this interaction, we account for the possibility that the indirect effect of gender on math performance is mediated differently depending on the students' positions within the network. The resulting network mediation model can be expressed as follows: 
\begin{equation}\label{mediation1}
    \begin{split}
        E(M_k \mid x, c) &= a_{0k} + a_{1k} \,x + a_{2k} \, c, \\
        E(Y \mid x, \mathbf{m}, c) &= b_0 + b_1 \, x + \sum_{k} b_{2k} \, m_k + \sum_{k} b_{3k} \,x \, m_k + b_4 \, c,
    \end{split}
\end{equation}
where $x$ is the binary indicator for gender \textcolor{black}{(female $=1$, male $=0$)}, $m_k$ represents the latent positions of the students, and $c$ is the confounder indicating proficiency based on the CST score threshold. {\color{black}Specifically, $\mathbf{m}_i=(z_{i1},\ldots,z_{id},w_{i1},\ldots,w_{id})$ contains the $2d$ sender and receiver coordinates for student $i$. Thus, $k=1,\ldots,2d$, with $d=2$ for EXP and $d=3$ for EOI.} The model coefficients capture distinct aspects of the gender-performance relationship: $a_{1k}$ quantifies gender's influence on network positioning, $b_{2k}$ measures how network positions affect mathematics performance, and $b_{3k}$ captures the predictor-mediator interaction between gender and network position, revealing how network-mediated effects vary by gender. Additionally, $a_{2k}$ and $b_4$ account for the influence of prior mathematical proficiency on network positions and post-test performance, respectively. $a_{0k}$ and $b_0$ are intercept terms for the network mediation model.

This model enables the decomposition of the total effect (TE) into two distinct components: the natural direct effect (NDE) and the natural indirect effect (NIE). The NDE represents the component of the effect of gender ($X$) on math performance ($Y$) that is independent of the mediators, while the NIE captures the effect transmitted through the mediating network structure. These components constitute the total effect (TE), providing a comprehensive measure of the impact of gender on math performance. The natural direct effect, the natural indirect effect, and the total effect are given by:
\begin{equation}\label{mediation2}
    \begin{split}
        NDE &= \left(b_1 + \sum_{k} b_{3k}\left(a_{0k} + a_{1k} \,x^{*} + a_{2k} \, c\right)\right)(x-x^{*}), \\
        NIE &= \left(\sum_{k} \left( b_{2k} \, a_{1k} +  b_{3k}\,  a_{1k} \, x \right) \right)(x-x^{*}), \\
        TE &= NDE + NIE.
    \end{split}
\end{equation}
These effects can be calculated from the regression coefficients estimated with the equation above. Our analysis examines how distinct types of classroom interactions mediate gender-based differences in mathematics performance by investigating two separate networks: the EXP network and the EOI network. For both networks, we incorporate students' prior mathematics achievement as a confounder to account for preexisting academic differences. Through the application of our network mediation model to these distinct interaction networks, we aim to investigate whether the mediating effects of classroom network structures vary systematically based on interaction type.

\subsubsection{Statistical Estimation Procedures}

\paragraph*{Stage 1: Latent Position Estimation}
We take a two-stage estimation approach for data analysis: (1) AMEN model estimation; and (2) network mediation model estimation. In the first stage, we estimate the proposed AMEN model to embed the observed student-interaction network into a latent space. The prior distributions of the AMEN model parameters are specified as follows:
\begin{equation}\label{prior}
\begin{split}
    \alpha_i &\sim N(0, \sigma_{\alpha}^2), \quad \sigma_{\alpha}^2 > 0, \quad i = 1, \dots, N\\
    \beta_j &\sim N(0, \sigma_{\beta}^2), \quad \sigma_{\beta}^2 > 0, \quad j = 1, \dots, N\\
    z_i &\sim N_d(0, \sigma_z^2I_d), \quad \sigma_{z}^2 > 0, \quad i = 1, \dots, N \\
    w_j &\sim N_d(0, \sigma_w^2I_d), \quad \sigma_{w}^2 > 0, \quad j = 1, \dots, N \\
    \log r &\sim N(\mu_r, \sigma_r^2), \quad \mu_r \in \mathbb{R}, \quad \sigma_r^2 > 0
\end{split}
\end{equation}

\noindent where \(\sigma_{\alpha}^2\), \(\sigma_{\beta}^2\), \(\sigma_{z}^2\), \(\sigma_{w}^2\), and \(\sigma_r^2\) are fixed prior variance terms and \(I_d\) is the \(d \times d\) identity matrix. \textcolor{black}{The latent dimension was selected using the lowest mean BIC, yielding \(d=2\) for EXP and \(d=3\) for EOI (detailed results are reported in Section \ref{supp-sec:amen} of the Supplementary Material).} Posterior samples are drawn via Markov chain Monte Carlo (MCMC), using the Metropolis-adjusted Langevin algorithm (MALA) for latent positions and a Metropolis-Hastings-within-Gibbs sampler for the remaining parameters. Because the multiplicative term \(z_i^T w_j\) is invariant under orthogonal transformations, we applied Procrustes matching \citep{gower1975generalized} to align posterior samples to a common reference configuration, resolving the rotational non-identifiability inherent in latent space models.

\paragraph*{Stage 2: Mediation Analysis}

In the second stage, we estimate a network mediation model. Specifically, the latent positions of the estimated AMEN model serve as mediators in the regression framework outlined in Equation (\ref{mediation2}). The network mediation model is estimated through Bayesian regression, with prior distributions specified as: 
\begin{equation*} 
\Theta \sim N(\boldsymbol{0}, I), \quad \epsilon_i \sim N(0, \sigma^2), \quad \sigma^2 \sim \text{Inverse-gamma}(0.001, 0.001), 
\end{equation*} 
\noindent where $\Theta \in \{a_{01}, \dots, a_{2k}, b_0, b_1, b_{21} \dots, b_{3k}, b_4\}$ represents all the regression coefficients in Equation (\ref{mediation1}), $\epsilon_i$ represents the error term in the regression model, and $\sigma^2$ is the variance of the error term. The network mediation model is estimated using the MCMCpack package \citep{martin2011mcmcpack} in {\tt R}. The estimated regression coefficients are then used to calculate the natural direct effect (NDE) and the natural indirect effect (NIE), as defined in Equation (\ref{mediation2}). 

To assess robustness, we ran 10 independent MCMC chains and monitored convergence using standard convergence diagnostics. Because different chains may converge to rotationally equivalent but numerically distinct solutions for the latent positions, we aligned all posterior samples to a common reference: the chain with the lowest BIC value, using its Maximum A Posteriori (MAP) estimate as the alignment target.

\subsubsection{Results of Latent Position Estimation}\label{sec:amen_results}

\begin{figure}[!htbp]
    \centering
    \begin{subfigure}[b]{\textwidth} 
        \centering
        \includegraphics[width=.90\textwidth]{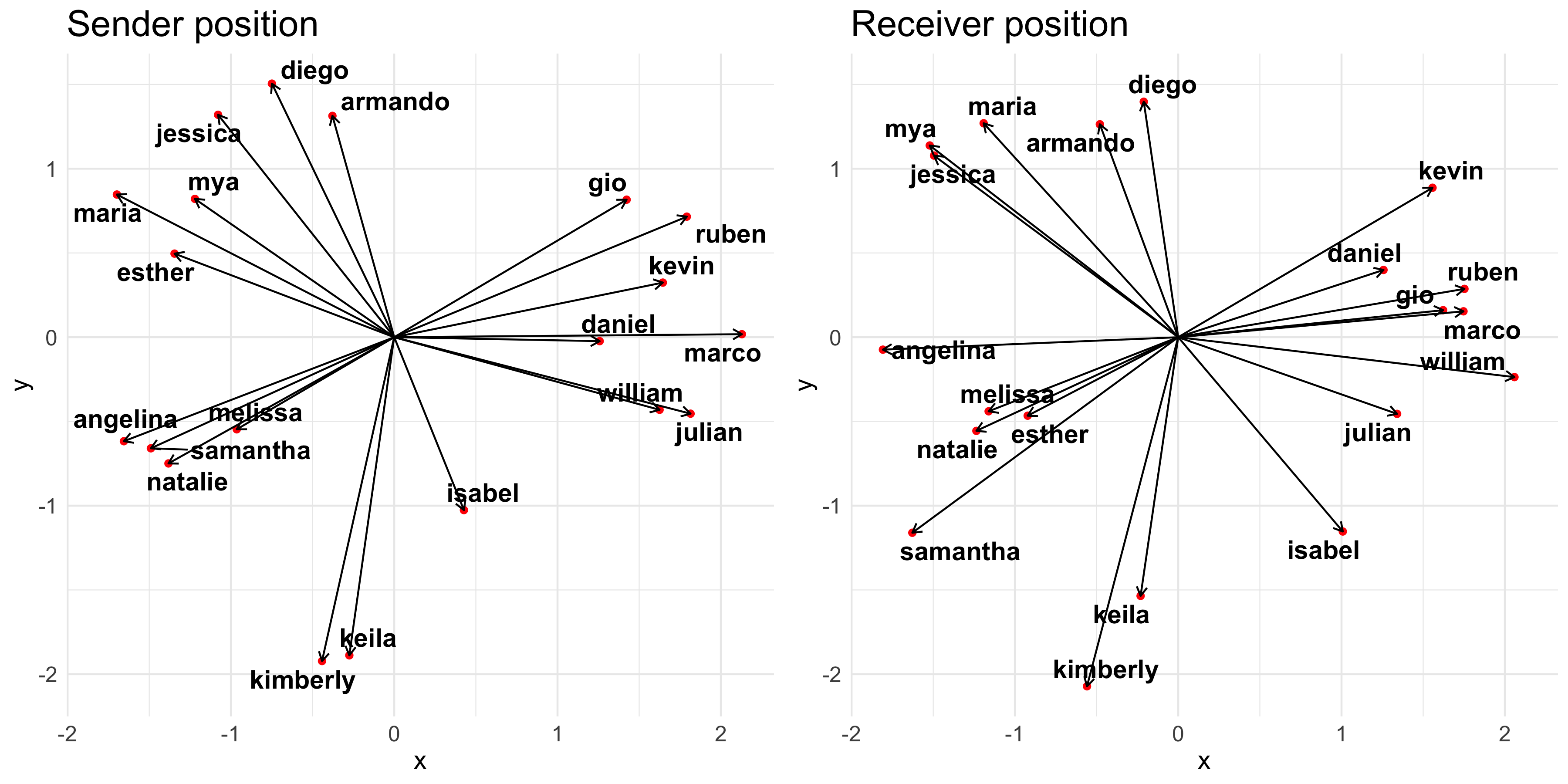}
        \caption{EXP (Explain Own Idea) Network}
        \label{fig:latentexp}
    \end{subfigure}
    
    \vspace{.3\baselineskip}

    \begin{subfigure}[b]{\textwidth} 
        \centering
        \makebox[\textwidth][c]{\includegraphics[width=.95\textwidth]{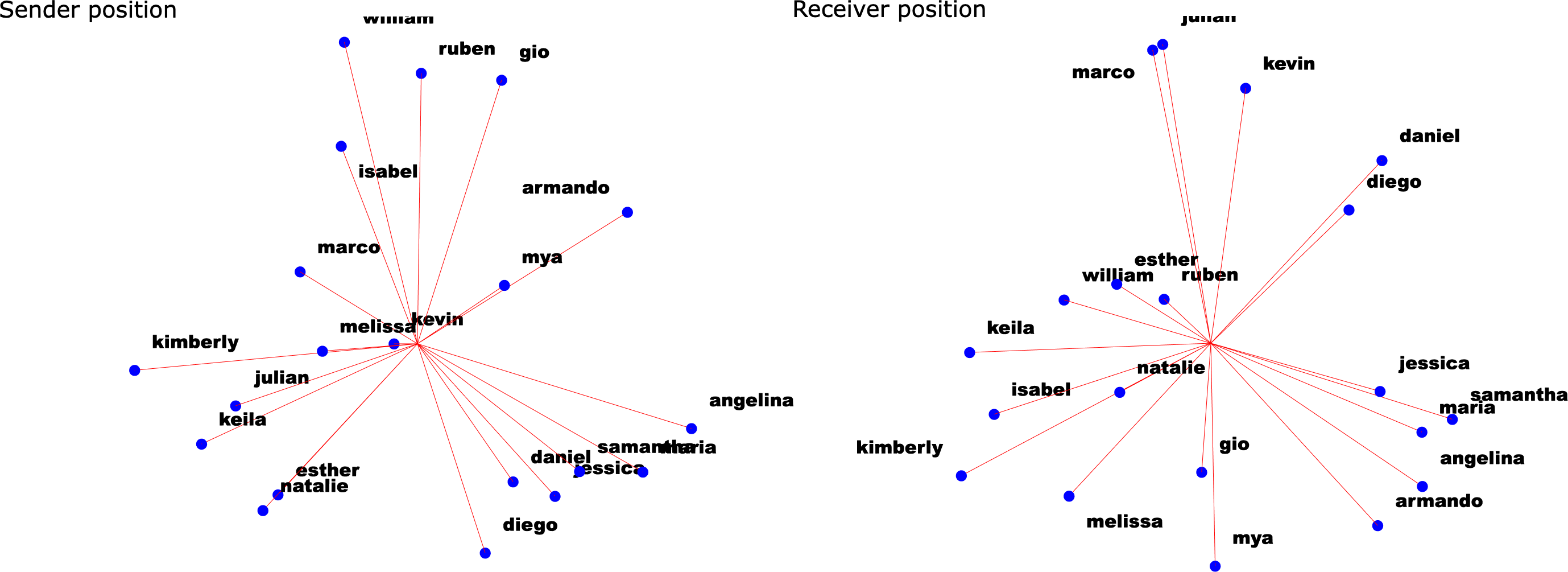}}
        \caption{EOI (Engage Others' Idea) Network}
        \label{fig:latenteoi}
    \end{subfigure}
    \caption{Comparison of Latent Positions for EXP and EOI Networks. Each node is represented as a point in a two-dimensional space; the EOI network ($d=3$) is projected onto two dimensions for visualization. The direction from the origin indicates similarity between nodes, and the distance from the origin indicates influence within the network.}
    \label{fig:latentall}
\end{figure}

\paragraph*{Interpretation of Latent Positions} 

Figures \ref{fig:latentexp} and \ref{fig:latenteoi} illustrate the latent positions of students within the EXP and EOI networks. \textcolor{black}{Each student's position is represented in two dimensions for EXP and three dimensions for EOI.} The angle between the student vectors indicates their interaction similarity; smaller angles represent greater similarity in interaction patterns. Additionally, the magnitude of the vector represents each student's relative influence within the network structure.

The latent space representations recover and extend the structural roles previously identified through centrality measures, while additionally revealing cluster structure and bridging relationships that individual centrality metrics cannot capture. In the EXP network, for instance, Gio occupies a position intermediate between two distinct clusters---one comprising Diego and Armando, and the other encompassing Julian, William, and Marco---consistent with his high betweenness centrality score and confirming his role as a primary intermediary. Similarly, Isabel is positioned so as to connect the Keila--Kimberly--Esther cluster with the Julian--William--Marco subgroup, performing a complementary bridging function.


In the EOI network, Isabel serves as a bridge connecting two distinct student groups: the Keila and Kimberly group and the Gio and Julian group. These patterns demonstrate a key advantage of latent space modeling over centrality analysis: the latent positions simultaneously encode directionality, interaction intensity, and higher-order relational structure within a single low-dimensional representation, providing a richer characterization of the social dynamics that will serve as mediators in the subsequent analysis.

\subsubsection{Results of Network Mediation Analysis}

Our investigation extends to examining how interaction patterns are associated with students' mathematical achievement through network mediation analysis. This analytical approach specifically evaluates how network structures mediate the relationship between gender and mathematics learning outcomes, quantifying the three key components: NIE, NDE, and TE. 

\begin{table}[H]
    \centering
    \begin{tabular}{lccccl}
\hline
  & Post.M & Post.SD & 2.5\% & 97.5\% & \\
\hline
NIE & \textcolor{black}{0.9885 (0.0198)} & \textcolor{black}{0.8674 (0.0076)} & \textcolor{black}{-0.4871 (0.0135)} & \textcolor{black}{2.8864 (0.0276)} & \\
NDE & \textcolor{black}{-1.4497 (0.0247)} & \textcolor{black}{1.1498 (0.0046)} & \textcolor{black}{-3.9154 (0.0299)} & \textcolor{black}{0.686 (0.0217)} & \\
TE & \textcolor{black}{-0.4612 (0.007)} & \textcolor{black}{0.7037 (0.0044)} & \textcolor{black}{-1.8414 (0.0076)} & \textcolor{black}{0.9971 (0.0197)} & \\
\hline
\end{tabular}
    \caption{Network Mediation Model Results for EXP Network. \textcolor{black}{Each entry is the mean of ten chain-specific summaries, with the between-chain standard deviation in parentheses.}}
    \label{tab:netmedexpc}
\end{table}

{\color{black}For the EXP network, the posterior means are 0.9885 for NIE, -1.4497 for NDE, and -0.4612 for TE. With female coded as 1 and male as 0, the positive NIE and negative NDE have opposing directions. The EXP-network-mediated component is positive for the female--male contrast, whereas the direct component is negative. Numerically, the indirect component partially offsets the negative direct component and yields a smaller negative TE. The posterior-position sensitivity analysis in the Supplementary Material shows the same predominant sign pattern across retained draws. However, all three 95\% credible intervals in the accompanying table include zero. The directional pattern is therefore descriptive and is not evidence of a nonzero effect or a compensatory mechanism.}

\begin{table}[H]
    \centering
    \begin{tabular}{lccccl}
\hline
  & Post.M & Post.SD & 2.5\% & 97.5\% & \\
\hline
NIE & \textcolor{black}{0.1782 (0.0458)} & \textcolor{black}{0.8347 (0.0085)} & \textcolor{black}{-1.5495 (0.0407)} & \textcolor{black}{1.7871 (0.0449)} & \\
NDE & \textcolor{black}{-0.6303 (0.0639)} & \textcolor{black}{1.001 (0.0082)} & \textcolor{black}{-2.7135 (0.0759)} & \textcolor{black}{1.2721 (0.0457)} & \\
TE & \textcolor{black}{-0.452 (0.0224)} & \textcolor{black}{0.591 (0.0091)} & \textcolor{black}{-1.6672 (0.0317)} & \textcolor{black}{0.7127 (0.012)} & \\
\hline
\end{tabular}
    \caption{Network Mediation Model Results for EOI Network. \textcolor{black}{Each entry is the mean of ten chain-specific summaries, with the between-chain standard deviation in parentheses.}}
    \label{tab:netmedeoic}
\end{table}

{\color{black}For the EOI network (Table \ref{tab:netmedeoic}), the posterior means are 0.1782 for NIE, -0.6303 for NDE, and -0.4520 for TE. Each 95\% credible interval includes zero, so the analysis does not support a nonzero direct or network-mediated effect in this illustrative application.}

\subsection{Transparency and Openness}
\textcolor{black}{An earlier version of this work was presented orally at IASC-ARS 2025 on December 3, 2025, and is available as a preprint at arXiv:2501.18912 (\url{https://arxiv.org/abs/2501.18912}). No part of this work has been published in a peer-reviewed publication.} We report all data preprocessing steps, utterance exclusions, classification models, and evaluation measures used in this study. \textcolor{black}{All analysis code, including the Python scripts for LLM-based classification and the R scripts for network construction and network analysis, is publicly available} at \url{https://github.com/gwangheekim/conversation-network-analysis.git}. The original classroom interaction data cannot be shared publicly due to participant privacy protections and ethical restrictions. \textcolor{black}{Commercial API requests transmitted only the transcript excerpts required for classification, consisting of utterance text, local conversational context, and first-name speaker labels of the form shown in the example tables. Surnames, full names, raw recordings, and other direct identifiers were not transmitted. OpenRouter documents its default logging practice and provider-specific retention information, while the applicable OpenAI, Anthropic, and Google commercial API policies state that submitted API content is not used for model training without the applicable customer authorization and describe provider-specific safety-monitoring or retention conditions \citep{openrouter2026faq,openrouter2026providerlogging,openai2026datacontrols,anthropic2026commercialtraining,google2026geminiterms}.} All analyses were conducted using Python, version 3.10, and R, version 4.5.1. This \textcolor{black}{study's} methodological design and analysis plan were not preregistered.

\section{Discussion}

In this study, we proposed a systematic methodology for constructing social networks from conversational data using ensemble prompt engineering. The prompt engineering approach enables automated classification of interaction patterns without requiring extensive pre-labeled datasets, substantially improving upon traditional manual labeling methods in both efficiency and scalability. \textcolor{black}{Our validation study assesses these advantages through external-criterion accuracy on an independently annotated corpus and a bounded comparison against human-coded episodes, rather than through inter-model consistency alone.} The ensemble approach produces robust labeling results while systematically identifying uncertain utterances through Shannon entropy-based uncertainty quantification, enabling targeted allocation of human review effort. The classified interactions are then encoded as directed weighted networks and projected onto latent spaces via the AMEN model with a negative binomial specification, yielding low-dimensional representations of the social structures that underlie observed interaction patterns.

Our contribution extends beyond traditional approaches through the integration of ensemble-based prompt engineering with \textcolor{black}{complementary validation evidence} and uncertainty quantification to construct \textcolor{black}{conversational networks}. To our knowledge, this combined approach with systematic uncertainty quantification has not been applied in dialogue analysis, representing a \textcolor{black}{methodological integration} that addresses both classification reliability challenges and the limitations of utterance-level analysis through the construction of relational structures.

To demonstrate the utility of networks constructed through this methodology, we presented two illustrative applications: network centrality analysis for descriptive characterization and network mediation analysis for statistical modeling. Our methodological integration demonstrates the potential of combining prompt engineering with network analysis to understand how conversational network structures are associated with outcomes of interest, with our mediation analysis being exploratory in nature and aiming to identify patterns and associations rather than establish causal relationships.

The proposed methodology has several limitations that suggest directions for future research. First, \textcolor{black}{the validation of our ensemble approach was conducted on the classroom corpus together with TalkMoves as a second related educational-discourse corpus and the generalizability of the observed reliability levels across other conversational contexts, languages or coding frameworks remains to be established.} Future work should evaluate this framework on diverse corpora---including clinical interviews, therapy sessions, and organizational team communications---to determine whether the reliability patterns observed here are robust across domains. Advancing beyond current approaches through model fine-tuning with high-quality labeled datasets would further enhance classification reliability in such diverse settings.

{\color{black}Second, the equal-weight ensemble offers a conservative aggregation rule when no model can be assumed to be uniformly superior, but it does not propagate classification uncertainty into the constructed edges or subsequent network model. Future work will extend the framework to incorporate measurement error throughout the utterance-to-network analysis.}

Third, current approaches rely heavily on human judgment for prompt creation, optimization, and selection, introducing potential variability and bias that can compromise analytical consistency across different research contexts. A particularly critical advancement involves developing automated prompt generation methodologies that reduce human dependency in the analytical pipeline. Automating these prompt development processes would enhance both operational efficiency and methodological consistency, transforming LLM-based classification into a more scalable and standardized tool for interaction analysis across research domains.

Finally, an important avenue for future research involves exploring temporal variations in conversational interaction networks. Time-varying network approaches could effectively capture the evolution of interaction patterns throughout the observation period, revealing how conversational dynamics change and develop over time. Such temporal analysis would illuminate whether the mediating effects of interaction networks vary across different phases of the process under study, potentially yielding more nuanced insights into the dynamic nature of conversational networks across diverse research contexts.

\section*{Acknowledgment}
\begin{sloppypar}
\noindent The authors thank the editor, associate editor, and reviewers for their constructive comments. The authors also thank Dr. Webb for providing the data and offering valuable insight for this study. This work was partially supported by the National Research Foundation of Korea [grant number NRF-2021S1A3A2A03088949, RS-2023-00217705, RS-2024-00333701; Basic Science Research Program awarded to IHJ], [grant number NRF-2025S1A5C3A0200632411; awarded to IHJ and MJ], International Joint Research Grant by Yonsei Graduate School [awarded to GK],  and the ICAN (ICT Challenge and Advanced Network of HRD) support program [grant number RS-2023-00259934], supervised by the IITP (Institute of Information \& Communications Technology Planning \& Evaluation).  Correspondence should be addressed to Ick Hoon Jin, Department of Applied Statistics, Department of Statistics and Data Science, Yonsei University, Seoul, Republic of Korea. E-mail: ijin@yonsei.ac.kr.
\end{sloppypar}

\bibliographystyle{apalike}
\bibliography{referenceGH}

@article{amen,
  title   = {Additive and Multiplicative Effects Network Models},
  author  = {Hoff, Peter D},
  journal = {Statistical Science},
  volume  = {36},
  number  = {1},
  pages   = {34--50},
  year    = {2021}
}

@article{networkmed,
  title     = {Network mediation analysis using model-based eigenvalue decomposition},
  author    = {Che, Chang and Jin, Ick Hoon and Zhang, Zhiyong},
  journal   = {Structural Equation Modeling: A Multidisciplinary Journal},
  volume    = {28},
  number    = {1},
  pages     = {148--161},
  year      = {2021},
  publisher = {Taylor \& Francis}
}

@article{fleiss,
  title     = {Measuring nominal scale agreement among many raters},
  author    = {Fleiss, Joseph L},
  journal   = {Psychological bulletin},
  volume    = {76},
  number    = {5},
  pages     = {378--382},
  year      = {1971},
  publisher = {American Psychological Association}
}

@article{shannon,
  title     = {A mathematical theory of communication},
  author    = {Shannon, Claude E},
  journal   = {The Bell system technical journal},
  volume    = {27},
  number    = {3},
  pages     = {379--423},
  year      = {1948},
  publisher = {Nokia Bell Labs}
}

@article{cohen,
  title     = {A coefficient of agreement for nominal scales},
  author    = {Cohen, Jacob},
  journal   = {Educational and psychological measurement},
  volume    = {20},
  number    = {1},
  pages     = {37--46},
  year      = {1960},
  publisher = {Sage Publications Sage CA: Thousand Oaks, CA}
}

@article{cohenkappa,
  title     = {The measurement of observer agreement for categorical data},
  author    = {Landis, J Richard and Koch, Gary G},
  journal   = {biometrics},
  pages     = {159--174},
  volume    = {33},
  number    = {1},
  year      = {1977},
  publisher = {JSTOR}
}

@article{lsm,
  title     = {Latent space approaches to social network analysis},
  author    = {Hoff, Peter D and Raftery, Adrian E and Handcock, Mark S},
  journal   = {Journal of the american Statistical association},
  volume    = {97},
  number    = {460},
  pages     = {1090--1098},
  year      = {2002},
  publisher = {Taylor \& Francis}
}

@article{long2024evaluating,
  title     = {Evaluating large language models in analysing classroom dialogue},
  author    = {Long, Yun and Luo, Haifeng and Zhang, Yu},
  journal   = {npj Science of Learning},
  volume    = {9},
  number    = {1},
  pages     = {60},
  year      = {2024},
  publisher = {Nature Publishing Group UK London}
}

@inproceedings{ganesh2024prompting,
  title     = {Prompting as Panacea? A Case Study of In-Context Learning Performance for Qualitative Coding of Classroom Dialog},
  author    = {Ganesh, Ananya and Chandler, Chelsea and D'Mello, Sidney and Palmer, Martha and von der Wense, Katharina},
  booktitle = {Proceedings of the 17th International Conference on Educational Data Mining},
  pages     = {835--843},
  year      = {2024},
  doi       = {10.5281/zenodo.12729966},
  url       = {https://www.educationaldatamining.org/edm2024/proceedings/2024.EDM-posters.95/}
}

@inproceedings{yue2024large,
  title     = {Large language model cascades with mixture of thoughts representations for cost-efficient reasoning},
  author    = {Yue, Murong and Zhao, Jie and Zhang, Min and Du, Liang and Yao, Ziyu},
  booktitle = {ICLR 2024 Workshop on Reliable and Responsible Foundation Models},
  year      = {2024}
}

@article{chen2025harnessing,
  title   = {Harnessing Multiple Large Language Models: A Survey on LLM Ensemble},
  author  = {Chen, Zhijun and Li, Jingzheng and Chen, Pengpeng and Li, Zhuoran and Sun, Kai and Luo, Yuankai and Mao, Qianren and Yang, Dingqi and Sun, Hailong and Yu, Philip S},
  journal = {arXiv preprint arXiv:2502.18036},
  year    = {2025}
}

@inproceedings{jiang2023llmblender,
  title     = {{LLM-BLENDER}: Ensembling Large Language Models with Pairwise Ranking and Generative Fusion},
  author    = {Jiang, Dongfu and Ren, Xiang and Lin, Bill Yuchen},
  booktitle = {The 61st Annual Meeting Of The Association For Computational Linguistics},
  volume    = {1},
  year      = {2023},
  publisher = {Association for Computational Linguistics},
  url       = {https://aclanthology.org/2023.acl-long.792/},
  doi       = {10.18653/v1/2023.acl-long.792},
  pages     = {14165--14178}
}

@article{lu2024merge,
  title   = {Merge, Ensemble, and Cooperate! A Survey on Collaborative Strategies in the Era of Large Language Models},
  author  = {Lu, Jinliang and Pang, Ziliang and Xiao, Min and Zhu, Yaochen and Xia, Rui and Zhang, Jiajun},
  journal = {arXiv preprint arXiv:2407.06089},
  year    = {2024}
}

@article{hoff2005bilinear,
  title     = {Bilinear mixed-effects models for dyadic data},
  author    = {Hoff, Peter D},
  journal   = {Journal of the american Statistical association},
  volume    = {100},
  number    = {469},
  pages     = {286--295},
  year      = {2005},
  publisher = {Taylor \& Francis}
}

@article{franke2015student,
  title     = {Student engagement with others’ mathematical ideas: The role of teacher invitation and support moves},
  author    = {Franke, Megan L and Turrou, Angela C and Webb, Noreen M and Ing, Marsha and Wong, Jacqueline and Shin, Nami and Fernandez, Cecilia},
  journal   = {The Elementary School Journal},
  volume    = {116},
  number    = {1},
  pages     = {126--148},
  year      = {2015},
  publisher = {University of Chicago Press Chicago, IL}
}

@article{webb2014engaging,
  title     = {Engaging with others’ mathematical ideas: Interrelationships among student participation, teachers’ instructional practices, and learning},
  author    = {Webb, Noreen M and Franke, Megan L and Ing, Marsha and Wong, Jacqueline and Fernandez, Cecilia H and Shin, Nami and Turrou, Angela C},
  journal   = {International Journal of Educational Research},
  volume    = {63},
  pages     = {79--93},
  year      = {2014},
  publisher = {Elsevier}
}

@article{gardner2019classroom,
  title     = {Classroom interaction research: The state of the art},
  author    = {Gardner, Rod},
  journal   = {Research on language and social interaction},
  volume    = {52},
  number    = {3},
  pages     = {212--226},
  year      = {2019},
  publisher = {Taylor \& Francis}
}

@inproceedings{liu2017dadnn,
  title     = {Using context information for dialog act classification in DNN framework},
  author    = {Liu, Yang and Han, Kun and Tan, Zhao and Lei, Yun},
  booktitle = {Proceedings of the 2017 conference on empirical methods in natural language processing},
  pages     = {2170--2178},
  year      = {2017}
}

@article{song2021automatic,
  title     = {Automatic classification of semantic content of classroom dialogue},
  author    = {Song, Yu and Lei, Shunwei and Hao, Tianyong and Lan, Zixin and Ding, Ying},
  journal   = {Journal of Educational Computing Research},
  volume    = {59},
  number    = {3},
  pages     = {496--521},
  year      = {2021},
  publisher = {SAGE Publications Sage CA: Los Angeles, CA}
}

@article{liu2021social,
  title     = {Social network mediation analysis: A latent space approach},
  author    = {Liu, Haiyan and Jin, Ick Hoon and Zhang, Zhiyong and Yuan, Ying},
  journal   = {Psychometrika},
  volume    = {86},
  number    = {1},
  pages     = {272--298},
  year      = {2021},
  publisher = {Springer}
}

@article{webb2023learning,
  title     = {Learning through explaining and engaging with others’ mathematical ideas},
  author    = {Webb, Noreen M and Franke, Megan L and Johnson, Nicholas C and Ing, Marsha and Zimmerman, Joy},
  journal   = {Mathematical Thinking and Learning},
  volume    = {25},
  number    = {4},
  pages     = {438--464},
  year      = {2023},
  publisher = {Taylor \& Francis}
}

@article{valeri2013mediation,
  title     = {Mediation analysis allowing for exposure--mediator interactions and causal interpretation: theoretical assumptions and implementation with SAS and SPSS macros.},
  author    = {Valeri, Linda and VanderWeele, Tyler J},
  journal   = {Psychological methods},
  volume    = {18},
  number    = {2},
  pages     = {137--150},
  year      = {2013},
  publisher = {American Psychological Association}
}

@article{sewell2016latent,
  title     = {Latent space models for dynamic networks with weighted edges},
  author    = {Sewell, Daniel K and Chen, Yuguo},
  journal   = {Social Networks},
  volume    = {44},
  pages     = {105--116},
  year      = {2016},
  publisher = {Elsevier}
}

@article{gpt1,
  title   = {Improving language understanding by generative pre-training},
  author  = {Radford, Alec and Narasimhan, Karthik and Salimans, Tim and Sutskever, Ilya},
  journal = {In preprint},
  year    = {2018}
}

@article{gpt2,
  title   = {Language Models are Unsupervised Multitask Learners},
  author  = {Radford, Alec and Wu, Jeff and Child, Rewon and Luan, David and Amodei, Dario and Sutskever, Ilya},
  journal = {In preprint},
  year    = {2019}
}

@article{gpt3,
  title   = {Language Models are Few-Shot Learners},
  author  = {Tom B. Brown and Benjamin Mann and Nick Ryder and Melanie Subbiah and Jared Kaplan and Prafulla Dhariwal and Arvind Neelakantan and Pranav Shyam and Girish Sastry and Amanda Askell and Sandhini Agarwal and Ariel Herbert-Voss and Gretchen Krueger and Tom Henighan and Rewon Child and Aditya Ramesh and Daniel M. Ziegler and Jeffrey Wu and Clemens Winter and Christopher Hesse and Mark Chen and Eric Sigler and Mateusz Litwin and Scott Gray and Benjamin Chess and Jack Clark and Christopher Berner and Sam McCandlish and Alec Radford and Ilya Sutskever and Dario Amodei},
  year    = {2020},
  journal = {ArXiv preprint ArXiv:2005.14165}
}

@article{xie2021explanation,
  title   = {An explanation of in-context learning as implicit bayesian inference},
  author  = {Xie, Sang Michael and Raghunathan, Aditi and Liang, Percy and Ma, Tengyu},
  journal = {ArXiv preprint ArXiv:2111.02080},
  year    = {2021}
}

@inproceedings{min2022rethinking,
  title     = {Rethinking the role of demonstrations: What makes in-context learning work?},
  author    = {Min, Sewon and Lyu, Xinxi and Holtzman, Ari and Artetxe, Mikel and Lewis, Mike and Hajishirzi, Hannaneh and Zettlemoyer, Luke},
  booktitle = {Proceedings of the 2022 Conference on Empirical Methods in Natural Language Processing},
  year      = {2022},
  publisher = {Association for Computational Linguistics},
  url       = {https://aclanthology.org/2022.emnlp-main.759},
  doi       = {10.18653/v1/2022.emnlp-main.759},
  pages     = {11048--11064}
}

@article{liu2023pre,
  title     = {Pre-train, prompt, and predict: A systematic survey of prompting methods in natural language processing},
  author    = {Liu, Pengfei and Yuan, Weizhe and Fu, Jinlan and Jiang, Zhengbao and Hayashi, Hiroaki and Neubig, Graham},
  journal   = {ACM Computing Surveys},
  volume    = {55},
  number    = {9},
  pages     = {1--35},
  year      = {2023},
  publisher = {ACM New York, NY}
}

@inproceedings{dong2024survey,
  title     = {A Survey on In-context Learning},
  author    = {Dong, Qingxiu and Li, Lei and Dai, Damai and Zheng, Ce and Ma, Jingyuan and Li, Rui and Xia, Heming and Xu, Jingjing and Wu, Zhiyong and Chang, Baobao and others},
  booktitle = {Proceedings of the 2024 Conference on Empirical Methods in Natural Language Processing},
  pages     = {1107--1128},
  year      = {2024}
}

@article{chi1997quantifying,
  title     = {Quantifying qualitative analyses of verbal data: A practical guide},
  author    = {Chi, Michelene TH},
  journal   = {The journal of the learning sciences},
  volume    = {6},
  number    = {3},
  pages     = {271--315},
  year      = {1997},
  publisher = {Taylor \& Francis}
}

@article{de2006content,
  title     = {Content analysis schemes to analyze transcripts of online asynchronous discussion groups: A review},
  author    = {De Wever, Bram and Schellens, Tammy and Valcke, Martin and Van Keer, Hilde},
  journal   = {Computers \& education},
  volume    = {46},
  number    = {1},
  pages     = {6--28},
  year      = {2006},
  publisher = {Elsevier},
  doi       = {10.1016/j.compedu.2005.04.005}
}

@article{di2022networks,
  title     = {Networks as mediating variables: a Bayesian latent space approach},
  author    = {Di Maria, Chiara and Abbruzzo, Antonino and Lovison, Gianfranco},
  journal   = {Statistical Methods \& Applications},
  volume    = {31},
  number    = {4},
  pages     = {1015--1035},
  year      = {2022},
  publisher = {Springer}
}

@article{martin2011mcmcpack,
  title   = {MCMCpack: Markov Chain Monte Carlo in R},
  author  = {Martin, Andrew D and Quinn, Kevin M and Park, Jong Hee},
  journal = {Journal of Statistical Software},
  volume  = {42},
  number  = {9},
  pages   = {1--21},
  year    = {2011}
}

@article{gower1975generalized,
  title     = {Generalized procrustes analysis},
  author    = {Gower, John C},
  journal   = {Psychometrika},
  volume    = {40},
  pages     = {33--51},
  year      = {1975},
  publisher = {Springer}
}

@article{vanderweele2009conceptual,
  title     = {Conceptual issues concerning mediation, interventions and composition},
  author    = {VanderWeele, Tyler J and Vansteelandt, Stijn},
  journal   = {Statistics and its Interface},
  volume    = {2},
  number    = {4},
  pages     = {457--468},
  year      = {2009},
  publisher = {International Press of Boston}
}

@misc{starprogram,
  title  = {Understanding California's Standardized Testing and Reporting (STAR) Program},
  author = {EdData},
  year   = {2013},
  note   = {Accessed January 17, 2025. \url{https://www.ed-data.org/article/Understanding-California's-Standardized-Testing-and-Reporting-(STAR)-Program}}
}

@inproceedings{CASA,
  title     = {Dialogue act classification with context-aware self-attention},
  author    = {Raheja, Vipul and Tetreault, Joel},
  booktitle = {Proceedings of the 2019 Conference of the North {A}merican Chapter of the Association for Computational Linguistics: Human Language Technologies, Volume 1 (Long and Short Papers)},
  year      = {2019},
  publisher = {Association for Computational Linguistics},
  url       = {https://aclanthology.org/N19-1373/},
  doi       = {10.18653/v1/N19-1373},
  pages     = {3727--3733}
}

@techreport{pagerank,
  number      = {1999-66},
  author      = {Lawrence Page and Sergey Brin and Rajeev Motwani and Terry Winograd},
  title       = {The PageRank Citation Ranking: Bringing Order to the Web.},
  type        = {Technical Report},
  publisher   = {Stanford InfoLab},
  year        = {1999},
  institution = {Stanford InfoLab}
}

@article{hubauth,
  title     = {Hubs, authorities, and communities},
  author    = {Kleinberg, Jon M},
  journal   = {ACM computing surveys (CSUR)},
  volume    = {31},
  number    = {4es},
  pages     = {5--es},
  year      = {1999},
  publisher = {ACM New York, NY, USA}
}

@article{grunspan2014understanding,
  title     = {Understanding classrooms through social network analysis: A primer for social network analysis in education research},
  author    = {Grunspan, Daniel Z and Wiggins, Benjamin L and Goodreau, Steven M},
  journal   = {CBE—Life Sciences Education},
  volume    = {13},
  number    = {2},
  pages     = {167--178},
  year      = {2014},
  publisher = {Am Soc Cell Biol}
}

@article{bokhove2018exploring,
  title     = {Exploring classroom interaction with dynamic social network analysis},
  author    = {Bokhove, Christian},
  journal   = {International Journal of Research \& Method in Education},
  volume    = {41},
  number    = {1},
  pages     = {17--37},
  year      = {2018},
  publisher = {Taylor \& Francis}
}

@article{williams2019linking,
  title     = {Linking engagement and performance: The social network analysis perspective},
  author    = {Williams, Eric A and Zwolak, Justyna P and Dou, Remy and Brewe, Eric},
  journal   = {Physical review physics education research},
  volume    = {15},
  number    = {2},
  pages     = {020150},
  year      = {2019},
  publisher = {APS}
}

@article{lindberg2010new,
  title     = {New trends in gender and mathematics performance: a meta-analysis.},
  author    = {Lindberg, Sara M and Hyde, Janet Shibley and Petersen, Jennifer L and Linn, Marcia C},
  journal   = {Psychological bulletin},
  volume    = {136},
  number    = {6},
  pages     = {1123--1135},
  year      = {2010},
  publisher = {American Psychological Association}
}

@article{ghasemi2019gender,
  title     = {Gender, affect, and math: a cross-national meta-analysis of Trends in International Mathematics and Science Study 2015 outcomes},
  author    = {Ghasemi, Ehsan and Burley, Hansel},
  journal   = {Large-scale Assessments in Education},
  volume    = {7},
  number    = {1},
  pages     = {1--25},
  year      = {2019},
  publisher = {Springer}
}

@article{else2010cross,
  title     = {Cross-national patterns of gender differences in mathematics: a meta-analysis.},
  author    = {Else-Quest, Nicole M and Hyde, Janet Shibley and Linn, Marcia C},
  journal   = {Psychological bulletin},
  volume    = {136},
  number    = {1},
  pages     = {103--127},
  year      = {2010},
  publisher = {American Psychological Association}
}

@article{bonacich1972factoring,
  title     = {Factoring and weighting approaches to status scores and clique identification},
  author    = {Bonacich, Phillip},
  journal   = {Journal of mathematical sociology},
  volume    = {2},
  number    = {1},
  pages     = {113--120},
  year      = {1972},
  publisher = {Taylor \& Francis}
}

@article{freeman1977set,
  title   = {A set of measures of centrality based on betweenness},
  author  = {Linton C. Freeman},
  journal = {Sociometry},
  number  = {1},
  pages   = {35--41},
  year    = {1977},
  volume  = {40}
}

@article{bavelas1948mathematical,
  title     = {A mathematical model for group structures},
  author    = {Bavelas, Alex},
  journal   = {Human organization},
  volume    = {7},
  number    = {3},
  pages     = {16--30},
  year      = {1948},
  publisher = {Society for Applied Anthropology}
}

@article{sedova2019those,
  title     = {Do those who talk more learn more? The relationship between student classroom talk and student achievement},
  author    = {Sedova, Klara and Sedlacek, Martin and Svaricek, Roman and Majcik, Martin and Navratilova, Jana and Drexlerova, Anna and Kychler, Jakub and Salamounova, Zuzana},
  journal   = {Learning and instruction},
  volume    = {63},
  pages     = {101217},
  year      = {2019},
  publisher = {Elsevier}
}

@article{mercer2010analysis,
  title     = {The analysis of classroom talk: Methods and methodologies},
  author    = {Mercer, Neil},
  journal   = {British journal of educational psychology},
  volume    = {80},
  number    = {1},
  pages     = {1--14},
  year      = {2010},
  publisher = {Wiley},
  doi       = {10.1348/000709909X479853}
}

@article{baum1966statistical,
  title     = {Statistical inference for probabilistic functions of finite state Markov chains},
  author    = {Baum, Leonard E and Petrie, Ted},
  journal   = {The annals of mathematical statistics},
  volume    = {37},
  number    = {6},
  pages     = {1554--1563},
  year      = {1966},
  publisher = {JSTOR}
}

@article{imel2015computational,
  title     = {Computational psychotherapy research: Scaling up the evaluation of patient-provider interactions},
  author    = {Imel, Zac E and Steyvers, Mark and Atkins, David C},
  journal   = {Psychotherapy},
  volume    = {52},
  number    = {1},
  pages     = {19--30},
  year      = {2015},
  publisher = {American Psychological Association}
}

@article{abdou2025leveraging,
  title     = {Leveraging Large Language Models to Estimate Clinically Relevant Psychological Constructs in Psychotherapy Transcripts},
  author    = {Abdou, Mostafa and Sahi, Razia S and Hull, Thomas D and Nook, Erik C and Daw, Nathaniel D},
  journal   = {Computational Psychiatry},
  volume    = {9},
  number    = {1},
  pages     = {187--209},
  year      = {2025},
  publisher = {MIT Press}
}

@article{goldberg2020machine,
  title     = {Machine learning and natural language processing in psychotherapy research: Alliance as example use case},
  author    = {Goldberg, Simon B and Flemotomos, Nikolaos and Martinez, Victor R and Tanana, Michael J and Kuo, Pai-Ju and Pace, Brian T and Villram, Zac E and others},
  journal   = {Journal of Counseling Psychology},
  volume    = {67},
  number    = {4},
  pages     = {438--448},
  year      = {2020},
  publisher = {American Psychological Association}
}

@article{marlow2018team,
  title     = {Does team communication represent a one-size-fits-all approach?: A meta-analysis of team communication and performance},
  author    = {Marlow, Shannon L and Lacerenza, Christina N and Paoletti, Jayden and Burke, C Shawn and Salas, Eduardo},
  journal   = {Organizational Behavior and Human Decision Processes},
  volume    = {144},
  pages     = {145--170},
  year      = {2018},
  publisher = {Elsevier}
}

@article{howe2013classroom,
  title     = {Classroom dialogue: A systematic review across four decades of research},
  author    = {Howe, Christine and Abedin, Manzoorul},
  journal   = {Cambridge journal of education},
  volume    = {43},
  number    = {3},
  pages     = {325--356},
  year      = {2013},
  publisher = {Taylor \& Francis},
  doi       = {10.1080/0305764X.2013.786024}
}

@article{michaels2008accountable,
  title     = {Deliberative Discourse Idealized and Realized: Accountable Talk in the Classroom and in Civic Life},
  author    = {Michaels, Sarah and O'Connor, Catherine and Resnick, Lauren B.},
  journal   = {Studies in Philosophy and Education},
  volume    = {27},
  number    = {4},
  pages     = {283--297},
  year      = {2008},
  publisher = {Springer},
  doi       = {10.1007/s11217-007-9071-1}
}

@article{chi2014icap,
  title     = {The {ICAP} Framework: Linking Cognitive Engagement to Active Learning Outcomes},
  author    = {Chi, Michelene T. H. and Wylie, Ruth},
  journal   = {Educational Psychologist},
  volume    = {49},
  number    = {4},
  pages     = {219--243},
  year      = {2014},
  publisher = {Taylor \& Francis},
  doi       = {10.1080/00461520.2014.965823}
}

@article{hennessy2016seda,
  title     = {Developing a Coding Scheme for Analysing Classroom Dialogue across Educational Contexts},
  author    = {Hennessy, Sara and Rojas-Drummond, Sylvia and Higham, Rupert and M\'{a}rquez, Ana Mar\'{i}a and Maine, Fiona and R\'{i}os, Rosa Mar\'{i}a and Garc\'{i}a-Carri\'{o}n, Roc\'{i}o and Torreblanca, Omar and Barrera, Mar\'{i}a Jos\'{e}},
  journal   = {Learning, Culture and Social Interaction},
  volume    = {9},
  pages     = {16--44},
  year      = {2016},
  publisher = {Elsevier},
  doi       = {10.1016/j.lcsi.2015.12.001}
}

@article{mercer1996quality,
  title     = {The Quality of Talk in Children's Collaborative Activity in the Classroom},
  author    = {Mercer, Neil},
  journal   = {Learning and Instruction},
  volume    = {6},
  number    = {4},
  pages     = {359--377},
  year      = {1996},
  publisher = {Elsevier},
  doi       = {10.1016/S0959-4752(96)00021-7}
}

@inproceedings{suresh2022talkmoves,
  title     = {The {TalkMoves} Dataset: {K--12} Mathematics Lesson Transcripts Annotated for Teacher and Student Discursive Moves},
  author    = {Suresh, Abhijit and Jacobs, Jennifer and Harty, Charis and Perkoff, Margaret and Martin, James H. and Sumner, Tamara},
  booktitle = {Proceedings of the Thirteenth Language Resources and Evaluation Conference},
  pages     = {4654--4662},
  year      = {2022},
  address   = {Marseille, France},
  publisher = {European Language Resources Association},
  url       = {https://aclanthology.org/2022.lrec-1.497/}
}

@misc{openrouter2026providerlogging,
  author       = {{OpenRouter}},
  title        = {Provider Logging: Provider Data Retention Policies},
  year         = {2026},
  howpublished = {OpenRouter Documentation, \url{https://openrouter.ai/docs/guides/privacy/provider-logging/}},
  url          = {https://openrouter.ai/docs/guides/privacy/provider-logging/},
  note         = {Accessed August 13, 2026}
}

@misc{openrouter2026faq,
  author       = {{OpenRouter}},
  title        = {OpenRouter {FAQ}: Privacy and Data Logging},
  year         = {2026},
  howpublished = {OpenRouter Documentation, \url{https://openrouter.ai/docs/faq}},
  url          = {https://openrouter.ai/docs/faq},
  note         = {Accessed August 17, 2026}
}

@misc{openai2026datacontrols,
  author       = {{OpenAI}},
  title        = {Data Controls in the {OpenAI} Platform},
  year         = {2026},
  howpublished = {OpenAI API Documentation, \url{https://developers.openai.com/api/docs/guides/your-data}},
  url          = {https://developers.openai.com/api/docs/guides/your-data},
  note         = {Accessed August 17, 2026}
}

@misc{anthropic2026commercialtraining,
  author       = {{Anthropic}},
  title        = {Is My Data Used for Model Training?},
  year         = {2026},
  howpublished = {Anthropic Privacy Center, \url{https://privacy.claude.com/en/articles/7996868-is-my-data-used-for-model-training}},
  url          = {https://privacy.claude.com/en/articles/7996868-is-my-data-used-for-model-training},
  note         = {Accessed August 17, 2026}
}

@misc{google2026geminiterms,
  author       = {{Google}},
  title        = {{Gemini API} Additional Terms of Service},
  year         = {2026},
  howpublished = {Google AI for Developers, \url{https://ai.google.dev/gemini-api/terms}},
  url          = {https://ai.google.dev/gemini-api/terms},
  note         = {Accessed August 17, 2026}
}

@article{woolley2010evidence,
  title     = {Evidence for a collective intelligence factor in the performance of human groups},
  author    = {Woolley, Anita Williams and Chabris, Christopher F and Pentland, Alex and Hashmi, Nada and Malone, Thomas W},
  journal   = {Science},
  volume    = {330},
  number    = {6004},
  pages     = {686--688},
  year      = {2010},
  publisher = {American Association for the Advancement of Science}
}

@article{bruun2019network,
  title     = {Network analysis and qualitative discourse analysis of a classroom group discussion},
  author    = {Bruun, Jesper and Lindahl, Mats and Linder, Cedric},
  journal   = {International journal of research \& method in education},
  volume    = {42},
  number    = {3},
  pages     = {317--339},
  year      = {2019},
  publisher = {Taylor \& Francis}
}

@article{liyanage2021student,
  title     = {Student discourse networks and instructor facilitation in process oriented guided inquiry physical chemistry classes},
  author    = {Liyanage, Dilhara and Lo, Stanley M and Hunnicutt, Sally S},
  journal   = {Chemistry Education Research and Practice},
  volume    = {22},
  number    = {1},
  pages     = {93--104},
  year      = {2021},
  publisher = {Royal Society of Chemistry}
}

@article{dawson2008study,
  title     = {A study of the relationship between student social networks and sense of community},
  author    = {Dawson, Shane},
  journal   = {Journal of educational technology \& society},
  volume    = {11},
  number    = {3},
  pages     = {224--238},
  year      = {2008},
  publisher = {JSTOR}
}

@misc{anthropic2026sonnet5,
  author       = {{Anthropic}},
  title        = {Claude Sonnet 5 System Card},
  year         = {2026},
  url          = {https://www-cdn.anthropic.com/73ad94ca3c0502e75e46637cc62c8bd9532a7f2c/Claude%20Sonnet%205%20System%20Card.pdf},
  note         = {Accessed 2026-08-16}
}

@misc{openai2026gpt56,
  author       = {{OpenAI}},
  title        = {{GPT-5.6}: Frontier Intelligence That Scales with Your Ambition},
  year         = {2026},
  url          = {https://openai.com/index/gpt-5-6/},
  note         = {Accessed 2026-08-16}
}

@misc{googledeepmind2026gemini38,
  author       = {{Google DeepMind}},
  title        = {Gemini 3.8 Flash},
  year         = {2026},
  url          = {https://deepmind.google/models/gemini/flash/},
  note         = {Accessed 2026-09-07}
}

@article{gemmateam2026gemma4,
  author       = {{Gemma Team}},
  title        = {Gemma 4 Technical Report},
  journal      = {arXiv preprint arXiv:2607.02770},
  year         = {2026},
  url          = {https://arxiv.org/abs/2607.02770}
}

@misc{qwenteam2026qwen36,
  author       = {{Qwen Team}},
  title        = {{Qwen3.6-27B}: Flagship-Level Coding in a 27B Dense Model},
  year         = {2026},
  month        = apr,
  url          = {https://qwen.ai/blog?id=qwen3.6-27b},
  note         = {Accessed 2026-08-16}
}

\newpage
\begin{abstract}
\noindent \textbf{Translational Abstract}\\
Conversational data, such as classroom discussions, therapy sessions, and team meetings, contain rich information about how people interact, but analyzing them has traditionally required extensive manual effort by trained researchers. This study presents a practical approach for \textcolor{black}{assisting} this process using large language models. \textcolor{black}{Five} models independently classify each utterance in a conversation, and \textcolor{black}{majority voting combines their labels. Cases where models disagree can be prioritized for review without treating agreement as evidence of accuracy.} The classified conversations are then used to construct \textcolor{black}{directed} interaction networks that reveal who communicates with whom and how meaningfully they engage. We demonstrate the approach with elementary mathematics classroom data, \textcolor{black}{illustrating} how interaction networks can identify which students play central roles in discussion and \textcolor{black}{support an exploratory analysis of whether} patterns of classroom interaction help explain gender differences in mathematics achievement. This methodology provides educators and researchers with a scalable tool for systematically examining conversational dynamics \textcolor{black}{in educational dialogue. Applications to other settings require domain-specific coding definitions, network-construction rules, and validation.}
\end{abstract}
\end{document}